\documentclass[aps,pr
,preprint
,showpacs
,groupedaddress
,superscriptaddress
,nobibnotes
]
{revtex4-1}
\usepackage{amsmath,amsthm,amssymb}
\usepackage{array}
\usepackage{subfigure}
\usepackage{graphicx}
\usepackage{fancyhdr}
\usepackage{extarrows}
\usepackage{enumerate}
\usepackage{epstopdf}
\usepackage{bm}
\usepackage[colorlinks,
          linkcolor=black,
            citecolor=black,
            urlcolor=blue
           ]{hyperref}
\usepackage{natbib}
\usepackage{comment}

\usepackage{xr}
\usepackage{tabularx}

\usepackage[normalem]{ulem}

\usepackage{color}
\usepackage{soul}

\begin{document}
\title{Topology, Vorticity and Limit Cycle in a Stabilized Kuramoto-Sivashinsky Equation
}
\author{Yong-Cong Chen}
\email{chenyongcong@shu.edu.cn}
\affiliation{Shanghai Center for Quantitative Life Sciences \& Physics Department, Shanghai University, Shanghai 200444, China}
\author{Chunxiao Shi}
\affiliation{Shanghai Center for Quantitative Life Sciences \& Physics Department, Shanghai University, Shanghai 200444, China}
\author{J. M. Kosterlitz}
\email[Corresponding author: ]{j\_kosterlitz@brown.edu}
\affiliation{Shanghai Center for Quantitative Life Sciences \& Physics Department, Shanghai University, Shanghai 200444, China}
\affiliation{Permanent Address: Department of Physics, Brown University, Providence, Rhode Island 02912, USA}
\author{Xiaomei Zhu}
\affiliation{Shanghai Center for Quantitative Life Sciences \& Physics Department, Shanghai University, Shanghai 200444, China}
\author{Ping Ao}
\affiliation{Shanghai Center for Quantitative Life Sciences \& Physics Department, Shanghai University, Shanghai 200444, China}


\begin{abstract}
A noisy stabilized Kuramoto-Sivashinsky equation is analyzed by stochastic decomposition. For values of control parameter for which periodic stationary patterns exist, the dynamics can be decomposed into diffusive and transverse parts which act on a stochastic potential. The relative positions of stationary states in the stochastic global potential landscape can be obtained from the topology spanned by the low-lying eigenmodes which inter-connect them. Numerical simulations confirm the predicted landscape. The transverse component also predicts a universal class of vortex like circulations around fixed points. These drive nonlinear drifting and limit cycle motion of the underlying periodic structure in certain regions of parameter space. Our findings might be relevant in studies of other nonlinear systems such as deep learning neural networks.
\end{abstract}

\maketitle

\renewcommand{\vec}[1]{\mathbf{#1}}
\newcommand{\bvec}[1]{\mbox{\boldmath $#1$}}
\newcommand{\cmmnt}[1]{\ignorespaces}


\section{Introduction}

 Complex systems far from equilibrium can rarely be described by well-established potentials or thermodynamic functions
 \cite{Prigogine1977,San1991,Elder1992,Grossmann_Kosterlitz,Tribelsky1996,Costa_Kosterlitz,Liang2013,Dunkel2013,Slomka2017}.
 Real world problems such as the Navier-Stokes (NS) equation \cite{Jolly1990,Anderson2006,Cross2009,Dunkel2013,Slomka2017} and artificial deep neural networks (DNN) \cite{doi:10.1146/annurev-conmatphys-031119-050745,Saxe11537,chaudhari2018stochastic,Fenge2015617118} are examples of such systems. However, the questions if, how and under what circumstances proper stochastic potentials can be constructed for such systems have been addressed recently by Ao {\em et~al.} \cite{Ao_2004,Ao_Thouless, Ao_2008,Yuan_Lyapunov,Yuan_Exploring,Zhu2006} These authors suggest that a stochastic system can possess a Lyapunov functional which describes some fluctuation dissipation properties of the system. There are two fundamentally distinct parts of the dynamics, a diffusive and a transverse process, both operating on the potential. This decomposition is unique near stationary points and is determined by the stochastic structure. The transverse process can lead to vorticity without detailed balance \cite{Ao_Thouless}.

 The methodology can be extended to nonlinear partial differential equations (PDEs) where the dynamical variables are labelled by continuous spatial coordinate(s). In an earlier work \cite{Chen23227}, a noisy one-dimensional stabilized Kuramoto-Sivashinsky (SKS) equation \cite{Misbah1994,Brunet2007,Pradas2011} was used to demonstrate the application of this. The SKS equation is derived formally \cite{Jolly1990} from an NS equation and it can describe a variety of physical phenomena with bifurcation instabilities \cite{Malomed1984,Kevrekidis1990,Goldstein1991,Knobloch1995}. The PDE exhibits nonlinear stationary cellular structures with additional complications such as vacillating breathing (VB) oscillations \cite{Misbah1994}. The absence of a conventional potential function \cite{Kerszberg1983,Obeid_Kosterlitz,Cross2016,Saxena_Kosterlitz} makes it a useful system for such stochastic studies.

In the following, we first review our earlier work \cite{Chen23227} on how to obtain a global potential landscape from a topological web of fixed points interconnected by low-lying eigenmodes. This result is then verified by direct stochastic simulations. The transverse dynamics near the fixed points and the nonlinear evolution of these are explored. A universal class of vortex like circulations is found near a range of cellular structures. The amplitude of circulation can grow or shrink with time and this is resilient to random noise. In a VB mode, a growing oscillation together with the nonlinearity exhibits limit cycles which cause periodic phase drifting of the cells themselves. We discuss our findings and their significance as a systematic alternative to explore nonlinearities.

\section{Stochastic Decomposition}

The noisy SKS equation is a one-dimensional nonlinear stochastic PDE which is periodic under $x\rightarrow x+L$ \cite{Hyman1986,Christiansen1997,Lan2008},
\begin{eqnarray}
\label{SKS}
 \partial_t u(x, t) & = & -\hat{L}(x)\,u(x, t) + [\partial_x u(x, t)]^2 + \xi(x, t)\\
\label{linear0}
 \hat{L}(x) & = & [\alpha + \partial_x^2 + \partial_x^4]
\end{eqnarray}
where $\xi(x, t)$ is an additive external Gaussian noise with $\langle\xi(x, t)\rangle=0$ and
\begin{equation}\label{diffusion}
 \langle\xi(x, t)\xi(x', t')\rangle = 2\epsilon D(x, x')\delta(t-t').
\end{equation}
Here, $\epsilon$ is the noise strength and the diffusion matrix $D(x, x')$ is symmetric and semi positive definite.

Following the work of Ao \cite{Ao_2004} and subsequent studies \cite{Ao_Thouless,Ao_2008}, one can recast the equation into the form \cite{Ao_Thouless,Ao_2008},
\begin{eqnarray}\label{matrix}
 \partial_t u(x, t) =&-&\int \text{d}x' \,\left[D(x, x') + Q (x, x'; \{u(x,t)\})\right] \cr
 &\times& \frac{\delta}{\delta u(x')}\Phi(\{u(x,t)\}) + \xi(x, t).
 \end{eqnarray}
This can be understood as multiplication of infinite dimensional matrices. The multiplication of two matrices of continuous degrees of freedom is weighted by $\text{d}x$ and $\delta/\delta u(x')$, written below as $\partial_{\vec{u}}$, is the functional differentation of the global potential $\Phi(\{u(x,t)\})$. We adopt a convention in which a boldface symbol indicates a matrix or vector labelled by $x$, while the same symbol in normal face indicates the corresponding matrix element so that Eq.~(\ref{matrix}) becomes
\begin{eqnarray}\label{matrix0}
 \partial_t\, \vec{u}(t) & = & - [\vec{D} + \vec{Q}]\cdot{\partial}_{\vec{u}}\Phi[\vec{u}(t)] + \bvec{\xi}(t).
\end{eqnarray}
Here $\vec{u}(t)$ is the state vector with components labelled by $x$ and both the semi-positive definite $\vec{D}=\vec{D}^{\dag}$ and the anti-symmetric $\vec{Q}=-\vec{Q}^{\dag}$ are square matrices defined by Eq.~(\ref{matrix}). With this decomposition, $\Phi[\vec{u}]$ becomes a Lyapunov functional for Eq.~(\ref{SKS}) which characterizes the dynamical properties of the system \cite{Yuan_Lyapunov,Ao_2004,Ao_Thouless,Smelyanskiy1997,Zhu2006}.

\subsection{Equation for the Global Potential}

We now briefly summarize the main conclusions of \cite{Chen23227}. For homogeneous and spatially uncorrelated noise, we set $\vec{D} = \vec{I}$ with matrix elements
\begin{equation}\label{diffusion0}
 I(x, x') = \delta(x-x').
\end{equation}
Letting $\vec{L} = \vec{L}^{\dag}$ be the linear operator in Eq.~(\ref{linear0}) with
\begin{equation}\label{linear0_1}
 L(x, x') = \hat{L}(x)\delta(x - x'),
\end{equation}
the linear term on the right-hand side of Eq.~(\ref{SKS}) corresponds to $- {\partial}_{\vec{u}}\Phi_{0}[\vec{u}(t)] $ with
\begin{equation}\label{potential00}
\Phi_{0}[\vec{u}] = \frac{1}{2}\,\vec{u}^{\dag}\,\vec{L}\,\vec{u}.
\end{equation}
The nonlinear term is recovered by setting $\vec{Q} = \vec{G}$ where
\begin{equation}\label{transfer0}
 G(x,x';\{u(x)\}) = u_x(x)[\hat{L}^{-1}(x')\partial_{x'}\delta(x-x')].
\end{equation}
However, to make $\vec{Q}$ antisymmetric we must adjust $\Phi$ and these are related by \cite{Chen23227}
\begin{equation}\label{DeltaPhi}
\left[\vec{G} - \vec{Q}\right]\partial_{\vec{u}}\Phi_0 -\left[\vec{I} + \vec{Q}\right]\partial_{\vec{u}}[\Phi- \Phi_0] = 0.
\end{equation}

Eq.~(\ref{DeltaPhi}) can be solved formally by defining a force $\vec{F}$ as the gradient of the potential
\begin{equation}\label{DeltaQPhiF}
 \vec{F} = -\partial_{\vec{u}}\Phi = -\left[\vec{I} + \vec{Q}\right]^{-1}\left[\vec{I} + \vec{G}\right]\vec{L} \,\vec{u}
\end{equation}
which must have vanishing curl,
\begin{equation}\label{DeltaQ}
 \partial_{\vec{u}} \times \vec{F} \equiv \frac{\delta F(x', \{u\})}{\delta u(x)} - \frac{\delta F(x, \{u\})}{\delta u(x')} = 0.
\end{equation}
Eq.~(\ref{DeltaQ}) determines $\vec{Q}$ and ensures that $\Phi(\{u\})$ is a path independent integral over the field variables,
\begin{equation}\label{DeltaPhi1}
 \Phi(\{u\}) = -\int \text{d}x\left\{\int_{0}^{u(x)}{\cal D}v\, F(x; \{v\})\right\}.
\end{equation}
These formal results suggest strongly the existence of a global potential for the entire system, although the nonlinearity in Eq.~(\ref{DeltaQPhiF}) is a major obstacle to its construction.

\subsection{Near Stationary States}\label{sec3}

We carry out the same procedure starting from a nontrivial fixed point solution $a(x)$ of Eq.~({\ref{SKS})
\begin{eqnarray}\label{La}
\hat{L}(x)a(x)=[\partial_x a(x)]^{2}
\end{eqnarray}
where $\tilde{u}(x)=u(x)-a(x)$ is the deviation from $a(x)$. The linear part of Eq.~({\ref{SKS}) is obtained from a slightly different potential
\begin{eqnarray}\label{potential1}
\Phi_{0}(\vec{u}:\vec{a})&=&\Phi(\vec{a}) + \frac{1}{2}\,\vec{\tilde{u}}^{\dag}\,\vec{L}\,\vec{\tilde{u}},
\end{eqnarray}
and the nonlinear part by the replacement $\vec{G}\rightarrow\tilde{\vec{G}}$ in Eq.~(\ref{transfer0}) where
\begin{eqnarray}\label{transfer1}
\tilde{G}(x,x';\{u:a\})&=& \cr
[\tilde{u}_{x}(x)&+& 2a_{x}(x)]\hat{L}^{-1}(x')\partial_{x'}\delta(x-x'),
\end{eqnarray}
Note, when $\vec{u}=0=\tilde{\vec{u}}+\vec{a}$, $\tilde{\vec{G}}\neq 0$ (cf. Eq.~(\ref{transfer0})). It is convenient to define $\vec{A} \equiv \tilde{\vec{G}}\,\vec{L} = \vec{A}_{0} + \vec{A}_{1}$ where
\begin{eqnarray}\label{transfer_a}
A_{0}(x,x';\{a\}) & = & 2a_x(x)\partial_{x'}\delta(x-x') \cr
A_{1}(x,x';\{\tilde{u}\}) & = & \tilde{u}_{x}(x)\partial_{x'}\delta(x-x').
\end{eqnarray}
At a fixed point, $\vec{A}\rightarrow \vec{A}_{0}$ and expanding Eq.~(\ref{DeltaQPhiF}) in powers of $\tilde{u}$ we have
\begin{eqnarray}\label{DeltaQPhi1_a}
\vec{F}_{1} & = & -\vec{R}_{0}\,\tilde{\vec{u}} + O(\tilde{u}^2), \cr
\vec{R}_{0} & = & \left[\vec{I} + \vec{Q}_{0}\right]^{-1}\left[\vec{L} + \vec{A}_{0}\right].
\end{eqnarray}
Here the subscripts indicate orders in powers of $\tilde{u}(x)$.

We obtain an equation for $\vec{Q}_{0}$ by observing that $\partial_{\vec{u}}\times\vec{F}_{1} = 0 \Rightarrow \vec{R}_{0}=\vec{R}_{0}^{\dag}$ so that
\begin{eqnarray}\label{DeltaQ2_a}
[\vec{L}+\vec{A}_{0}]\vec{Q}_{0}+\vec{Q}_{0}[\vec{L}+\vec{A}_{0}^{\dag}]=\vec{A}_{0}-\vec{A}_{0}^{\dag}.
\end{eqnarray}
Eq.~(\ref{DeltaQ2_a}) is known as a continuous Lyapunov equation \cite{Mori2002,Jbilou2006,Hached2018} for which there exist efficient numerical algorithms \cite{Ao_Thouless,Chen23227}. From Eq.~(\ref{DeltaQPhi1_a}) the potential to ${\cal O}(\tilde{u}^{2})$ is
\begin{eqnarray}\label{DeltaPhi_a}
\Phi_{2}(\vec{u}: \vec{a}) & = & \frac{1}{2}\,\vec{\tilde{u}}^{\dag}\,\vec{R}_{0}\,\vec{\tilde{u}} + \Phi(\vec{a}).
\end{eqnarray}

\section{Topology and Global Landscape}

Knowing the potential near individual fixed points allows us explore the global properties of the system. When $L\rightarrow\infty$ and $ \alpha < 1/4$, the SKS equation has a continuous band of periodic stationary states \cite{Misbah1994,Brunet2007} and part of the band is stable. When $L<\infty$ the states can be labelled by the wave number $\kappa=2\pi k/L$ with integer $k$, centered around a critical wave number $\kappa_c = 1/\sqrt{2}$. However, in the presence of external noise some states are more stable than others which can be understood as a natural consequence of a global potential. In the following, we show how the potential differences between these fixed points can be inferred from the topology spanned by a network of interconnected fixed points. The analysis is supplemented by direct stochastic simulations.

\subsection{Potential Difference Between Stationary States}

If we extrapolate $\Phi_2$ of Eq.~(\ref{DeltaPhi_a}) to a neighboring fixed point $u(x) = b(x)$, the potential difference between them, assuming that a single valued potential exists, would be approximately $\Phi_2(\vec{b}: \vec{a})$ of Eq.~(\ref{DeltaPhi_a}). Since the same procedure applies in the opposite direction from $b(x)$ to $a(x)$, the potential difference should be
\begin{eqnarray}\label{DeltaPhi3}
\Delta\Phi_{ba}=\frac{1}{2}[\Phi_{2}(\vec{b}: \vec{a}) - \Phi_{2}(\vec{a}: \vec{b})]=\Phi(\vec{b})-\Phi(\vec{a}).
\end{eqnarray}
This approach can be refined by noticing that the entire set of fixed points forms an interconnected web \cite{Chen23227}. There is always a pair of dominant eigenmodes of $\vec{R}_{0}$ leaving from one state and flowing towards another state. These modes can be identified as having the largest amplitude with the wave number of the destination state, together with an eigenvalue with a vanishing real part. This novel topology suggests that Eq.~(\ref{DeltaPhi3}) should be confined to the subspace of the interconnected modes only so that the dominant contribution to the landscape is from the low-lying modes flowing between the nodes. Define $\vec{v}^{\sigma}_{ba}$ ($\sigma=\pm$) to be the eigenmodes of $\vec{R}_{0}$ at state $a$ flowing to state $b$ with eigenvalue $\lambda^{\sigma}_{ba}$. An improved version of Eq.~(\ref{DeltaPhi3}) is
\begin{eqnarray}\label{DeltaPhi4}
\Delta\Phi_{ba}\approx \sum_{\sigma=\pm}\frac{1}{4}\,\vec{c}^{\dag}\,[\vec{v}^{\sigma}_{ba}\lambda^{\sigma}_{ba}\vec{v}^{\sigma\dag}_{ba} -\vec{v}^{\sigma}_{ab}\lambda^{\sigma}_{ab}\vec{v}^{\sigma\dag}_{ab}]\,\vec{c},\;\;\;\vec{c}\equiv \vec{b} - \vec{a}.
\end{eqnarray}

Knowing the pairwise potential differences, one can map out the global potential difference between any two states by following a path between them. However, this potential difference is path dependent and, to make the result path independent as it must be, we include the whole set of pairs to obtain $\Phi(\kappa)$ as a function $\kappa$ by a least squares fit to a low-order polynomial. A more detailed discussion is in the supplementary information (SI) \cite{supplementary}. Also in the SI \cite{supplementary} we correct an error in our earlier work where there is an erroneous factor $h$ in the expression $(h\,a_{k-k'})$ in Eqs.~(34) and (35) of \cite{Chen23227}.

\subsection{Verification by Stochastic Simulations}

\begin{figure}[!ht]
\begin{tabular}{ccc}
\includegraphics[width=0.35\textwidth]{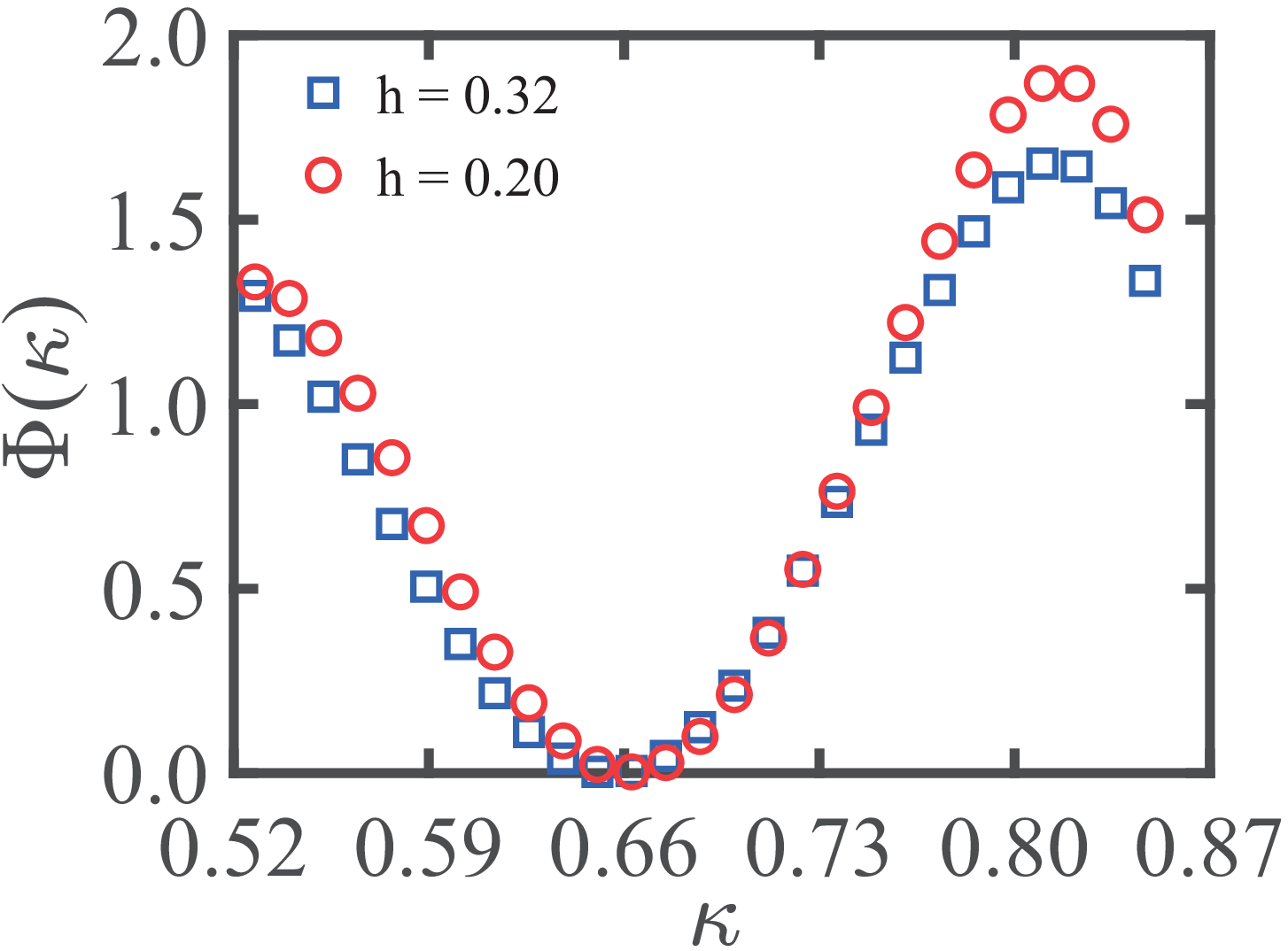} &
\includegraphics[width=0.35\textwidth]{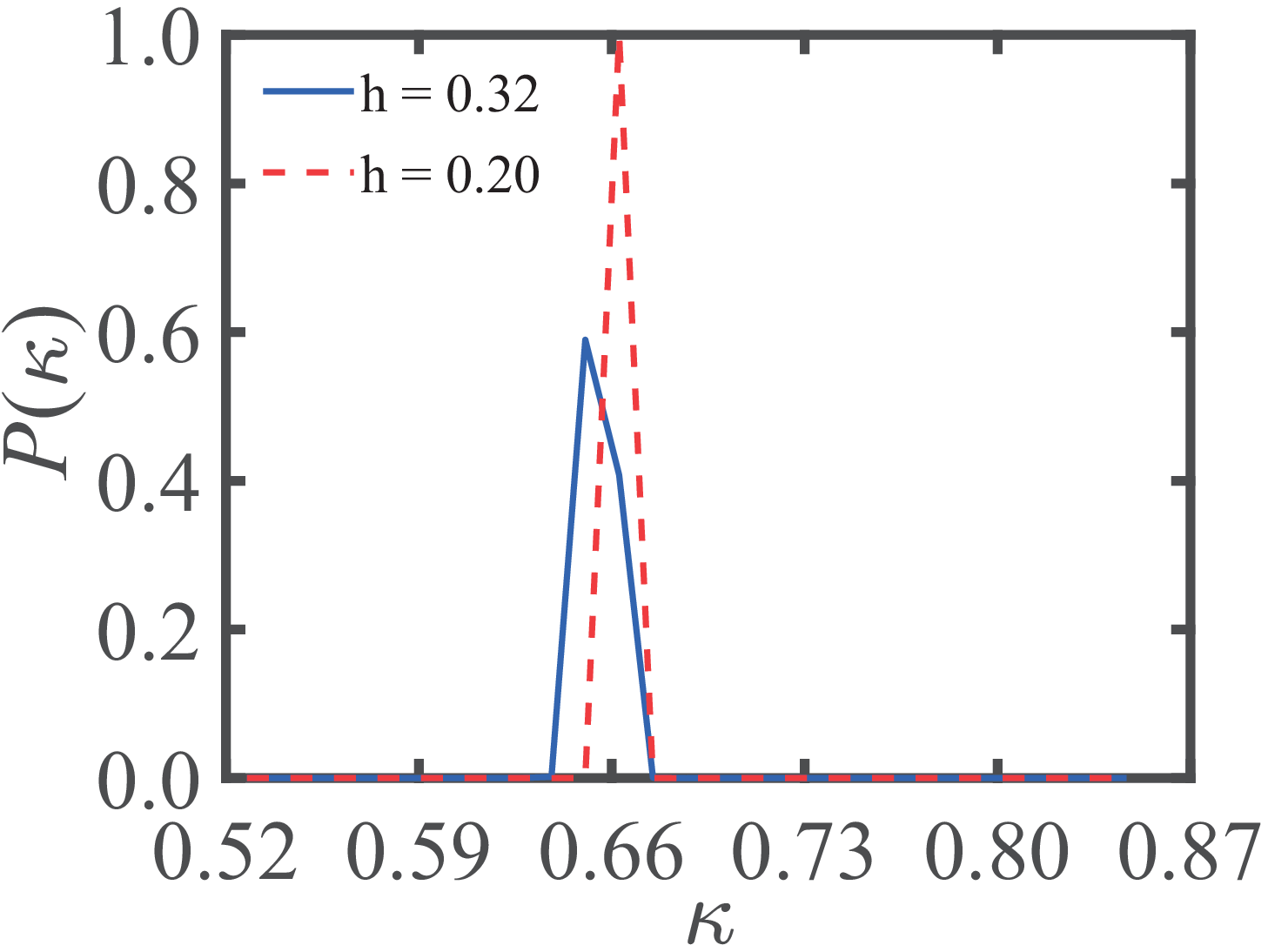} \\
(a) & (b) \\
\includegraphics[width=0.35\textwidth]{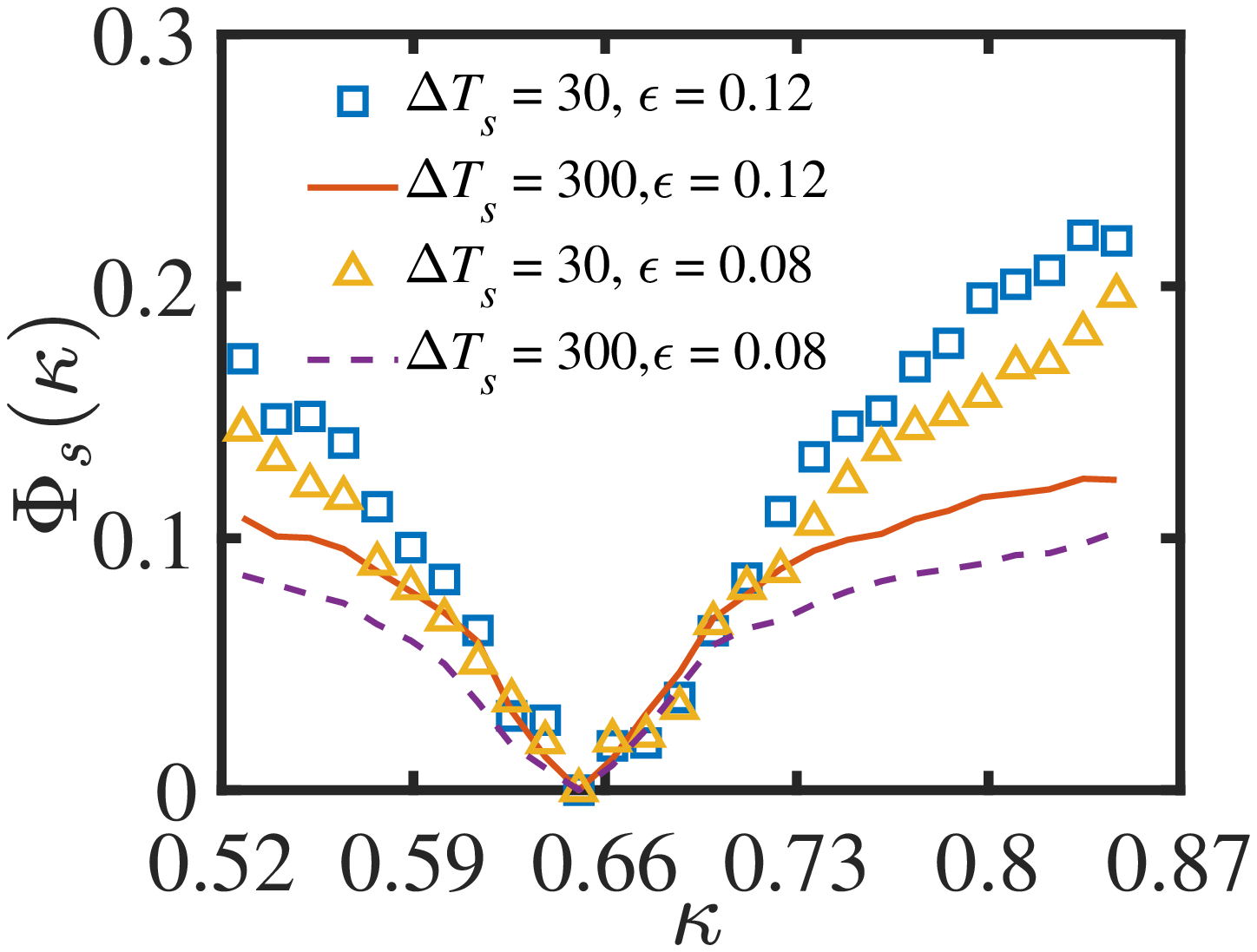} &
\includegraphics[width=0.35\textwidth]{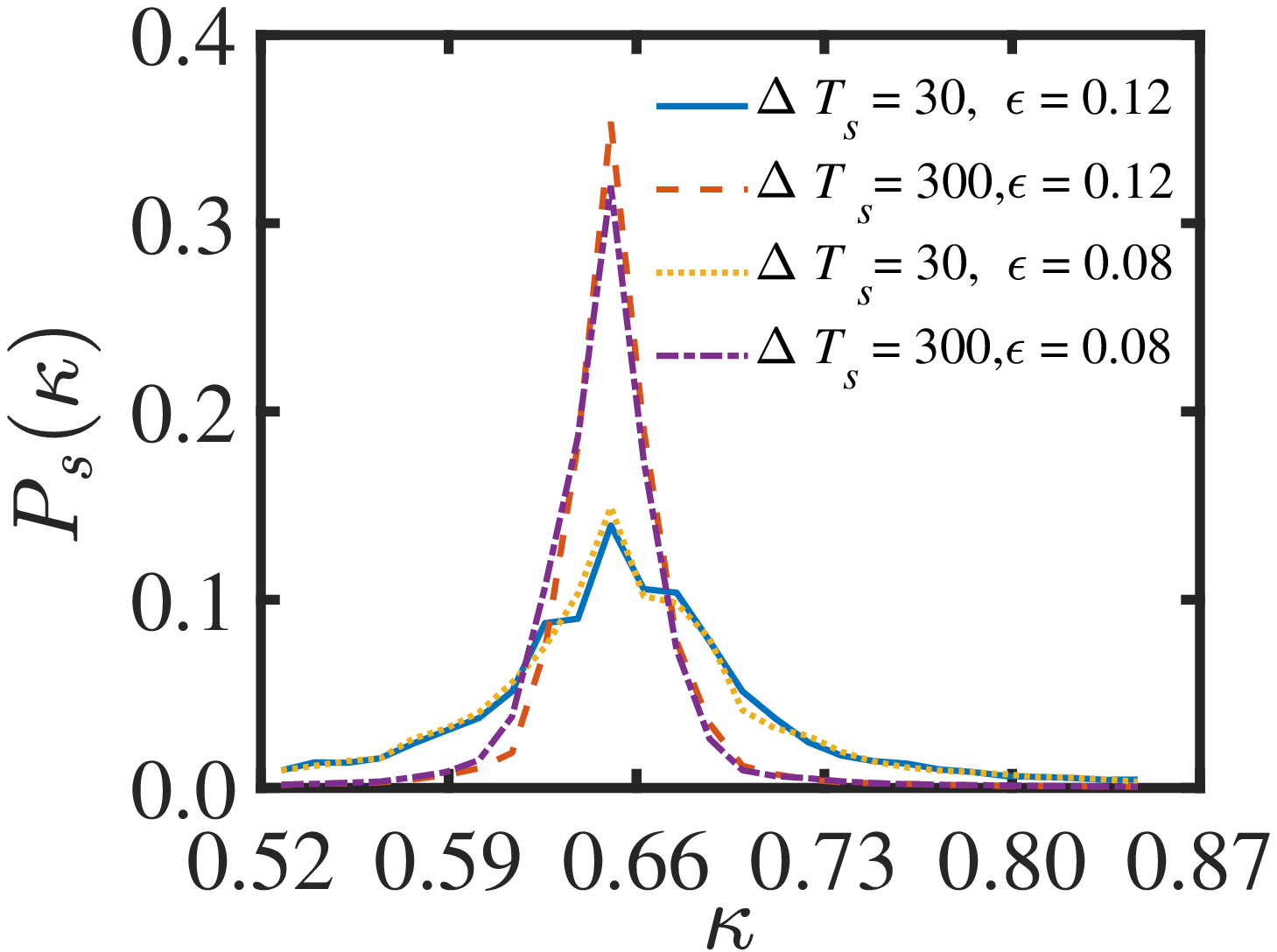} \\
(c) & (d) \\
\end{tabular}
\protect\caption{Global potentials (a), (c) and corresponding probability distributions (b), (d) for $L = 512$, $\alpha = 0.20$. (a) Global potentials $\Phi(\kappa)$ of Eq.~(\ref{DeltaPhi4}) from $4^{\rm th}$ order polynomial fits over the whole topology of stationary states for two grid spacings $h$. (b) Probability distributions $P(\kappa)$ using $\Phi(\kappa)$ of (a). (c), (d) Simulated potential $\Phi_s(\kappa)$ and distribution $P_s(\kappa)$ for $h = 0.32$ and several values of $\epsilon$ and $\Delta T_s$.}\label{Fig1}
\end{figure}

The global landscape $\Phi(\kappa)$ can be verified by comparing with $\Phi_{s}(\kappa)$ from direct stochastic simulations for the probability distribution $P(\kappa)$ in the presence of strong external noise with the algorithm of \cite{Saxena_Kosterlitz}. We expect $P(\kappa)$ is a Boltzmann-like distribution \cite{Ao_Thouless}, $P(\kappa) = P_{0}(\kappa)\exp[-\Phi(\kappa)/\epsilon]$ where $\epsilon$ is the noise strength of Eq.~(\ref{diffusion}) and $P_{0}(\kappa)$ is a slowly varying function of $\kappa$, although there is no rigorous proof of this. In a simulation with external stochastic noise, there is also the question of the meaning of occupying a stationary state $\kappa$.

Suppose the system is initially in some arbitrary state and the simulation is performed in the presence of external stochastic noise for some arbitrarily chosen time $t_{0}$. One can define the probability of being in the state $\kappa$ by the overlap of this state with the stationary solution of the noiseless SKS equation with wave number $\kappa$. A closely related method is to expand the simulated state at $t_{0}$ as a linear superposition of periodic solutions of the {\it noiseless} SKS equation and define its wave number as that of the periodic solution of the SKS equation of largest magnitude. Neither approach is satisfactory because neither accurately reproduces the theoretical potential $\Phi(\kappa)$. A third and better method is to switch off the noise at some sufficiently long time and then evolve the system in the absence of noise for a time $\Delta T_{s}$ to a stationary state of wave number $\kappa$. By repeating this many times, a simulated $P_{s}(\kappa)$ of a Boltzmann form is obtained with a simulated potential $\Phi_{s}(\kappa)$ which is a close match to the theoretical $\Phi(\kappa)$. However, the detailed shape of $P_{s}(\kappa)$ does depend on the time $\Delta T_{s}$ allowed for the chosen state to evolve to a stationary state. When an effective noise strength $\tilde{\epsilon} = \sqrt{\epsilon/\Delta T_{s}}$ is used to characterize the distribution $P_{s}(\kappa)$, we obtain a consistent $\Phi_{s}(\kappa)$ which is independent of the separate values of $\epsilon$ and $\Delta T_{s}$. The simulations agree reasonably well with the theoretical predictions up to an overall scale factor $\Phi(\kappa)/\Phi_{s}(\kappa)\sim 10$. Using $\alpha = 0.20$ as an example, a least squares polynomial fit and a stochastic simulation are compared in Fig.~\ref{Fig1}.} More simulation details can be found in the SI \cite{supplementary}.

\section{Vorticity near Fixed Points}

Another essential feature, which is a more distinct characteristic of the stochastic dynamics, is the transverse component described by the antisymmetric $\vec{Q}$ in Eq.~(\ref{matrix0}). When $\vec{Q}$ is large there is a large deviation from the gradient diffusion process. Vortex like circulation or ``vorticity'' can be a prominent feature of the dynamics. This can be explored near a steady state when $\vec{Q}\rightarrow \vec{Q}_{0}$ is essentially a constant matrix (the subscript $0$ and the overhead tilde on $\vec{u}$ are dropped in the following for simplicity).

\subsection{Oscillating Pair Decomposition}\label{subsection4a}

We are free to choose any convenient basis to represent the state vector. When $\vec{Q}$ is large, we choose the eigenvectors which partially diagonalize $\vec{Q}$ into a direct sum of pairs of $2\times 2$ antisymmetric matrices. Let $\bvec{q}_{i}=q_{i}\,(i\bvec{\sigma}_{y})$ where $q_{i}>0$ is the $i^{{\rm th}}$ eigenvalue and $\bvec{\sigma}_{y}$ is a Pauli matrix so that $\vec{Q} = \bvec{q}_{1}\oplus \bvec{q}_{2}\oplus\cdots \oplus\bvec{q}_{N/2}$. Denote the corresponding eigenvectors by $\vec{e}_{i\sigma}$ where $i=1,2, \dots, N/2$ and $\sigma=1, 2$ so that $\vec{e}^{\dag}_{i\sigma}\,\vec{Q}\,\vec{e}_{j\sigma'} = \delta_{ij}(\vec{q}_{i})_{\sigma\sigma'}$.

Following \cite{Ao_2008}, we define $\vec{S}+\vec{T} \equiv [\vec{I}+\vec{Q}]^{-1} $ so that $\vec{S}$ is a symmetric ``dissipative'' matrix and $\vec{T}$ is an antisymmetric ``transfer'' matrix. Now Eq.~(\ref{matrix0}) can be written as
\begin{eqnarray}\label{matrix1}
 [\vec{S} + \vec{T}]\,\partial_t \vec{u}(t) & = & - {\partial}_{\vec{u}}\Phi[\vec{u}(t)] + \bvec{\zeta}(t)
\end{eqnarray}
where the new ``canonical'' noise $\bvec{\zeta}(t) = [\vec{S}+\vec{T}]\,\bvec{\xi}(t)$ has zero mean and variance
\begin{equation}\label{diffusion1}
\langle\bvec{\zeta}(t)\bvec{\zeta}^{\dag}(t')\rangle = 2\epsilon \,\vec{S}\,\delta(t-t').
\end{equation}
Both $\vec{S}$ and $\vec{T}$ are diagonal in the same basis as $\vec{Q}$. Now, let $\vec{1}$ be the $2\times 2$ unit matrix so that, in the $i^{{\rm th}}$ subspace, $\vec{s}_{i} = s_{i}\vec{1}$ with $s_{i} = 1/(1+q_{i}^{2}) > 0$ and $\vec{t}_{i}$ is a $2\times 2$ antisymmetric matrix where $-(\bvec{t}_{i})_{12} =(\bvec{t}_{i})_{21}=t_{i} = q_{i}/(1+q_{i}^{2})$. When $q_{i}\gg 1$, all matrix elements are very small and $t_{i}/s_{i} \gg 1$. Since $\vec{S}$ relates dissipation to fluctuations by Eq.~(\ref{diffusion1}), a small $\vec{s}_{i}$ allows for oscillations of $\vec{u}_{i}$ in the $i^{{\rm th}}$ subspace by the transfer matrix $\vec{t}_{i}$ (cf. below). Note, when continuous matrices are discretized, the matrix element of $\vec{I}$ is not always $1$ but it can always be re-scaled so that this subtlety does not change the essence of our analysis.

The eigenstates can be labelled by $s_{1} \leq s_{2}\leq \dots \leq s_{N/2}$ and, in the $i^{{\rm th}}$ subspace, the lowest approximation to Eq.~(\ref{matrix1}) is
\begin{equation}\label{lc0}
(\vec{s}_{i}+\vec{t}_{i})\,\partial_{t}\vec{u}_{i}(t) = -\vec{r}_{ii}\,\vec{u}_{i}(t) + \bvec{\zeta}_{i}(t),
\end{equation}
where $(\vec{r}_{ij})_{\sigma\sigma'} = \vec{e}^{\dag}_{i\sigma}\,\vec{R}\,\vec{e}_{j\sigma'}$. When $\vec{s}_{i}\rightarrow 0$, the variance of the noise $\langle\bvec{\zeta}_{i}(t)\bvec{\zeta}^{\dag}_{i}(t')\rangle \rightarrow 0$ so that $\vec{u}_{i}$ of Eq.~(\ref{lc0}) oscillates with frequency $\omega_{i}\approx q_{i}\sqrt{\text{det}(\vec{r}_{ii})}$ when $\text{det}(\vec{r}_{ii}) > 0$. This oscillation either decays to a stable fixed point or grows away from an unstable fixed point. In either case, this creates vortex motion as discussed below.

\begin{figure}[!ht]
\begin{tabular}{ccc}
\includegraphics[width=0.30\textwidth]{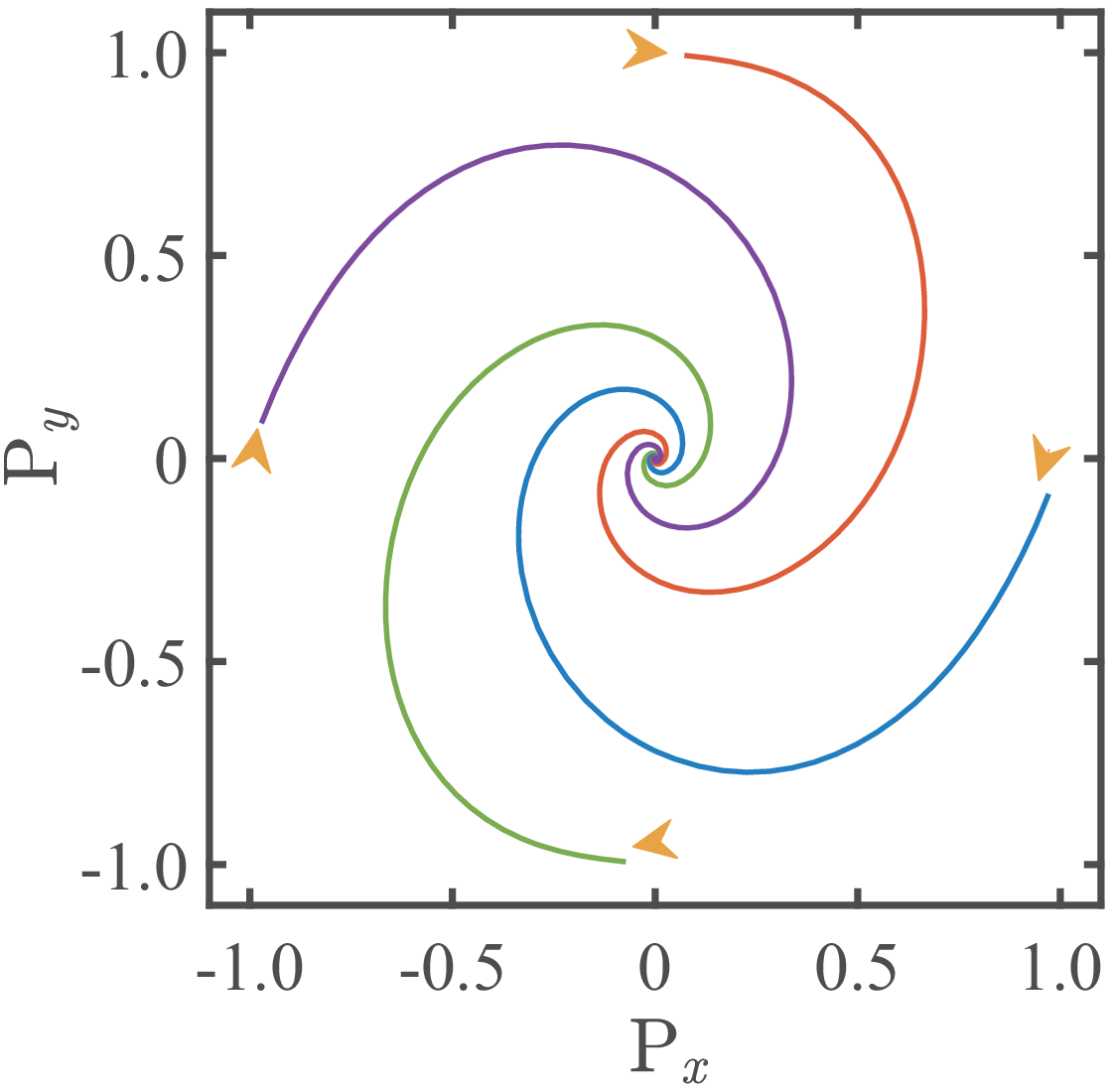} &
\includegraphics[width=0.30\textwidth]{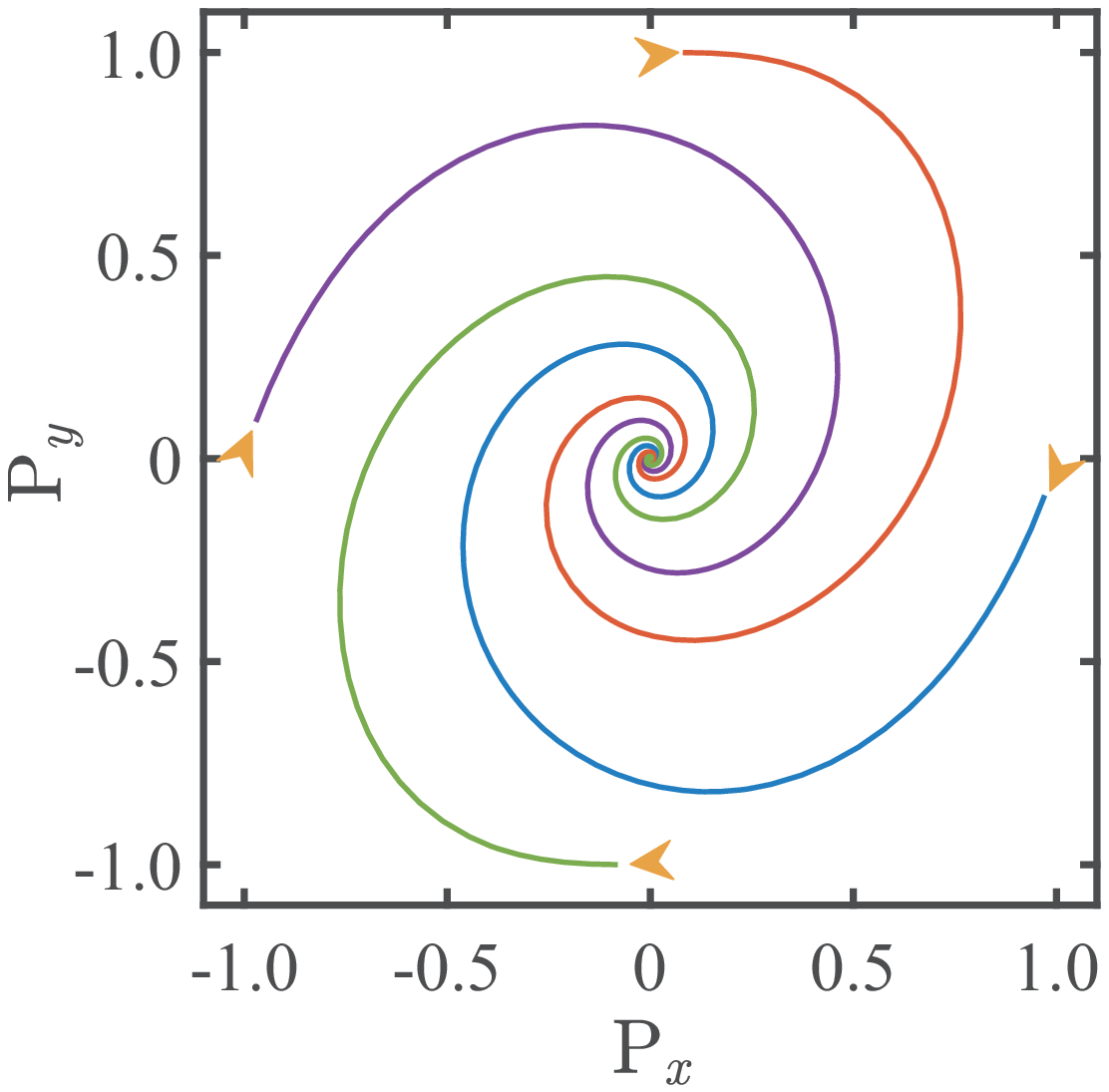} &
\includegraphics[width=0.30\textwidth]{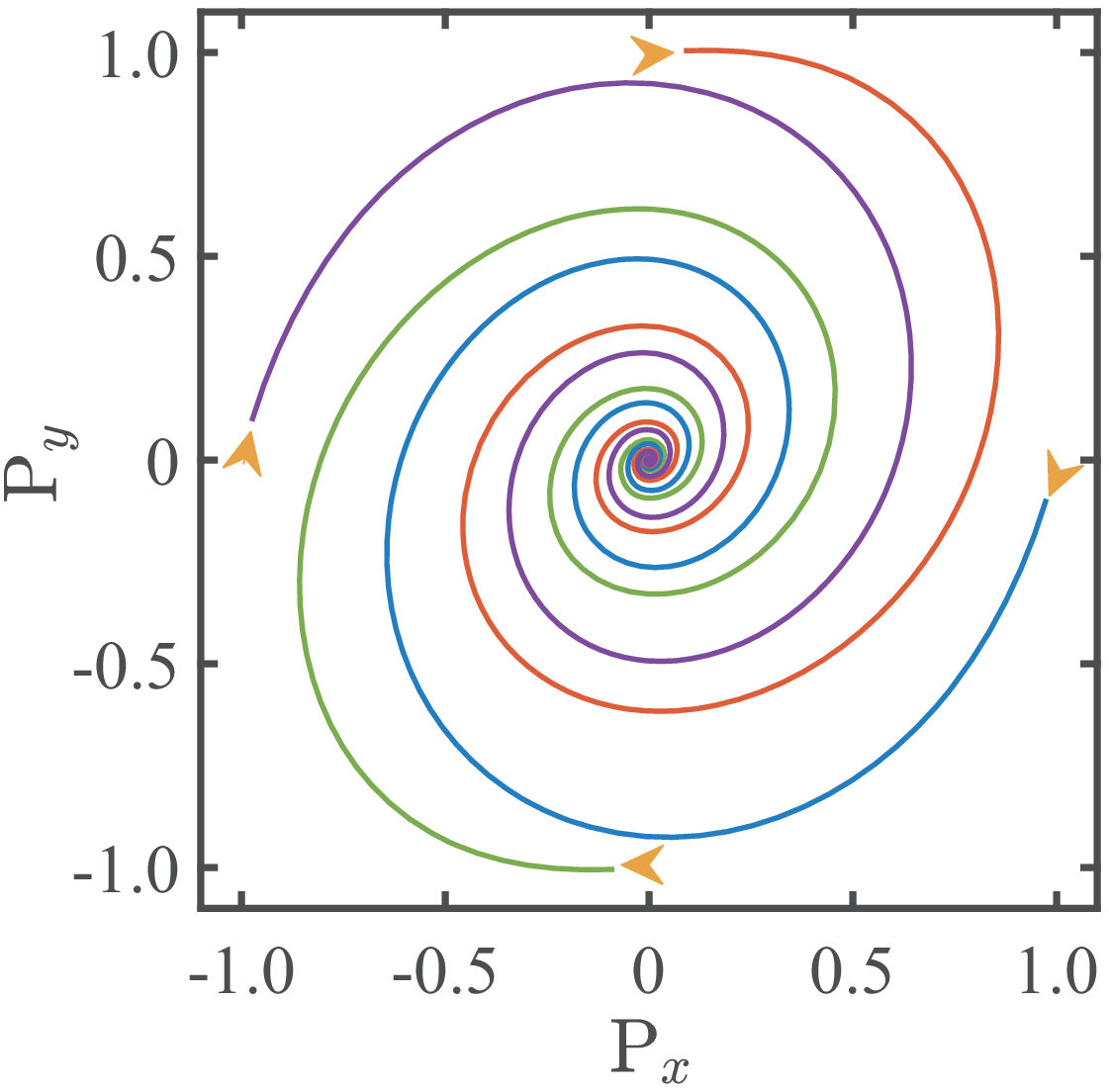} \\
(a) $\alpha = 0.17$ & (b) $\alpha = 0.15$ & (c) $\alpha = 0.12$ \\
\end{tabular}
\protect\caption{Vortex like motions in periodic structures with $L = 512$, $h = 0.32$, $\kappa = 0.6995$ for various $\alpha$. The system is initially in the $1^{\rm{st}}$ subspace of $\vec{Q}$ by Eq.~(\ref{lc0}). (a)-(c) Small deviations $P_{\sigma} = \vec{e}^{\dag}_{1\sigma}\cdot\tilde{\vec{u}}(t)$ ($\sigma=x,y$) from a stable state decay to zero for $\alpha = 0.17, 0.15, 0.12$ respectively.}\label{Fig2}
\end{figure}

A typical example of vortex motion near a steady state in the stable region is shown in Fig.~\ref{Fig2}, where $L = 512$, $h = 0.32$, wavenumber $\kappa = 0.6995$ and $\alpha = 0.17, 0.15, 0.12$. We choose to restrict the motion to the $1^{\rm{st}}$ subspace of $\vec{Q}$ from Eq.~(\ref{lc0}). Small initial deviations from the stationary state are chosen as the Fourier space eigenstates of $\vec{Q}$, $\tilde{\vec{u}}_{0} = \vec{e}_{1\sigma}$ ($\sigma = 1, 2$ or $x,y$ for convenience). These states evolve according to Eq.~\cmmnt{(\ref{SKSF})}(S10) in the SI \cite{supplementary}. The specific parameters chosen are: time step $\Delta t = 0.003$, number of iterations $10^6$ and data is recorded every $100^{{\rm th}}$ time step. The state vector is projected on to the $i^{{\rm th}}$ subspace by ${\rm{P}}_{\sigma}(t) = \vec{e}^{\dag}_{i\sigma}\cdot\tilde{\vec{u}}(t)$. More detailed discussion is found in the SI \cite{supplementary}.

\subsection{Overlap with exact eigenstates}

\begin{figure}[!ht]
\includegraphics[width=0.5\textwidth]{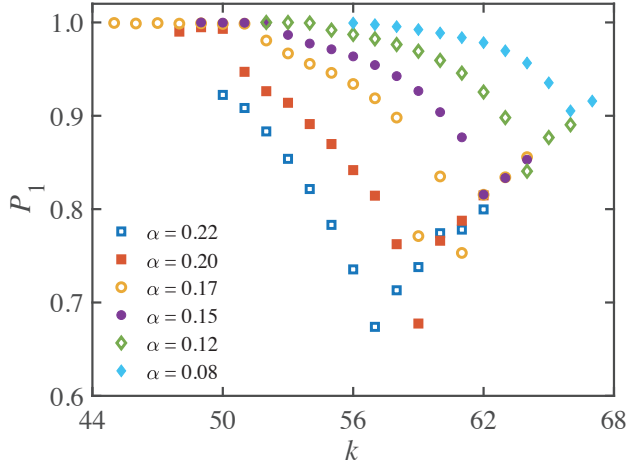}
\protect\caption{Overlap $P_{1}$ between two 4 dimensional degenerate subspaces, the $1^{\rm{st}}$ eigenstate $\vec{e}_{1\sigma}$ of $\vec{Q}$ and $\vec{V}_{j\sigma}$ ($\sigma = 1,2,3,4$) of $(\vec{D}+\vec{Q})\vec{R}$, varies on the Fourier component $k$ for different values of control parameter $\alpha$.}\label{Fig3}
\end{figure}

Here we investigate how accurately the two-state truncation represents the real many dimensional system as the control parameter $\alpha$ is reduced. There are higher order corrections to Eq.~(\ref{lc0}) from other pairs when $\bvec{r}_{ii}\gg\bvec{r}_{ij}\neq 0$ when the system is equivalent to a set of weakly coupled harmonic oscillators. A displacement $\vec{u}_{i}(t)$ drives the $j^{{\rm th}}$ pair by a force $\sim\vec{r}_{ji}\,\vec{u}_{i}(t)$ which adds to the right-hand side of Eq.~(\ref{lc0}) a perturbation $\vec{r}_{ij}\,\vec{u}_{j}(t)$.
Taking this into account, a second-order perturbation calculation, neglecting the random noise, yields
\begin{equation}\label{lc1}
(\vec{s}_{i}+\vec{t}_{i})\,\partial_{t}\vec{u}_{i}(t) = -\left[\vec{r}_{ii}-\sum_{j\neq i}\vec{r}_{ij}\,\frac{1}{(\vec{s}_{j}+\vec{t}_{j})\partial_{t}+\vec{r}_{jj}}\,\vec{r}_{ji}\right]\,\vec{u}_{i}(t).
\end{equation}
Writing ${\bf u}_{i}(t)={\bf u}_{i}(0)\,{\rm exp}(\lambda_{i}t)$, gives the secular equation
\begin{equation}\label{lc1}
\text{det}\left[(\vec{s}_{i}+\vec{t}_{i})\lambda_{i} + \vec{r}_{ii} - \sum_{j\neq i}\vec{r}_{ij}\,\frac{1}{(\vec{s}_{j}+\vec{t}_{j})\lambda_{i}+\vec{r}_{jj}}\,\vec{r}_{ji}\right] = 0,
\end{equation}
which can be evaluated iteratively. The real part of $\lambda_{i}$ is the damping or growth rate while the imaginary part, when it exists, gives the oscillation frequency $\omega_{i}$.

This approximation is in the right direction, but is not sufficient when quasi degenerate modes are involved. We can diagonalize numerically the $N\times N$ matrix $(\vec{D}+\vec{Q})\,\vec{R}$ in Eq.~(\ref{matrix0}) which yields all eigenvalues $\Lambda_{j\sigma}$ and eigenvectors $\vec{V}_{j\sigma}$, $(j=1,\dots,N/2)$. The overlap between the two spaces $\vec{e}_{i\sigma}$ and $\vec{V}_{j\sigma}$ can be obtained from the $2\times 2$ matrix $\vec{p}_{ij}$ with elements $(\vec{p}_{ij})_{\sigma\sigma'} = \vec{e}^{\dag}_{i\sigma}\vec{V}_{j\sigma'}$. An absolute measure of overlap is obtained from
\begin{equation}\label{overlap}
0\leq P_{ij}=\text{Tr}(\vec{p}_{ij}^{\dag}\vec{p}_{ij})/\text{Tr}(\vec{1}) \leq 1.
\end{equation}
An estimate of the overlap is obtained from $P_{i}={\rm max}_{j}(P_{ij})\equiv P_{ij_{m}}$ which identifies the correct eigenvalue as $\lambda_{i}=\Lambda_{j_{m}}$. $P_{i}$ is a measure of the isolation of the subspace from the larger environment and the larger $P_{i}$ is, the better is the two-state approximation to the dynamics near the fixed point. If two pairs of $\vec{e}_{i\sigma}$ and $\vec{V}_{j\sigma}$ are degenerate, it is convenient to compute the overlap between the two $4\times 4$ subspaces. Numerical results are shown in Fig.~\ref{Fig3}.

\subsection{Drifting of Steady States and Limit Cycles}

This general analysis can be applied to perturbations and vorticity about a periodic stationary state. Some of the analysis is most conveniently done in Fourier space but we return to real space to ensure that $\vec{\tilde{u}}$ is real. Algebraic and computation details can be found in the SI \cite{supplementary}.

The stochastic decomposition allows for a relatively simple identification of vortex modes and observation of their evolution in a nonlinear system. When $\text{Re}\,\lambda_{i} > 0$ the $i^{\text{th}}$ mode is unstable and its amplitude increases with time. Some modes are saturated by the nonlinearity and form a quasi limit cycle when their amplitude is sufficiently large. This behavior is seen clearly for values of control parameter region for which VB modes exist \cite{Misbah1994} (cf. below). Also, other interesting phenomena related to drifting of the periodic stationary states are seen.

In a VB mode, every cell oscillates out of phase by $\pi$ with its neighbors resulting in a quasi stationary periodic cellular structure which drifts uniformly in coordinate space. We find that this phenomenon can be attributed to the following generic pattern. Initially, the system is in a periodic state with a maximum at $x=0$. The eigenmodes of a small perturbation about this state are found to alternate between stable and unstable modes, when numbered from the smallest eigenvalue of the $\vec{S}$ matrix (with the minor complication that these modes are two-fold degenerate). We impose the $i^{\text{th}}$ unstable mode as an initial perturbation. The amplitude of this mode increases which causes uniform drifting of the quasi stationary periodic state. When the growth of this mode is saturated by the nonlinearity, it changes to a decaying mode which continues to drift. A careful analysis shows that the original mode is projected on to a mode which is stable relative to the new drifting stationary state. However, part of the amplitude also becomes a new unstable mode which begins to grow. The quasi steady state itself evolves back to its initial state, thus completing a limit cycle. We find that this pattern is quite robust against small external noise. More results are shown in the SI \cite{supplementary}.

\section{Discussions}

The topology of multiple inter connected fixed points in a global potential landscape subject to nonlinearity and random fluctuations is discussed in this research. The stochastic decomposition provides new insights into the dynamics near stationary points. From very general considerations, we predict the existence of vortex like limit cycles near stationary solutions when the dynamics has a significant transverse component (large $\vec{Q}$ in Eq.~(\ref{matrix0})). This prediction from theory agrees with numerical simulations. This explains and reproduces in detail the VB mode in the SKS equation. The limit cycles appear for certain values of the control parameter when the strength of the random fluctuations is sufficiently small.

These intriguing phenomena, which are generic in out of equilibrium nonlinear stochastic systems, may be useful for increasing our understanding of vorticity and turbulence in related systems. In addition to problems of natural origin, artificial ones such as DNN fall into this class, for example, the statistical mechanics of deep learning \cite{doi:10.1146/annurev-conmatphys-031119-050745} and pattern formation in semantic development \cite{Saxe11537} are very similar to the stochastic dynamics studied here. Even though the stochastic gradient descent in the learning process usually explicitly uses a cost function, a large anisotropy in the noise spectrum leads to a different canonical potential by the same decomposition used here, cf. \cite{chaudhari2018stochastic}. This results in limit cycles \cite{chaudhari2018stochastic} and an unusual inverse Einstein relation \cite{Fenge2015617118} near local minima. Further study, extension and use of the ideas and methods in this work seem to be worth further study.

\begin{acknowledgments}
This work was supported in part by the National Natural Science Foundation of China No. 16Z103060007 (PA). JMK thanks the Shanghai Center for Quantitative Life Sciences and Shanghai University for their hospitality while a portion of this work was begun.
\end{acknowledgments}


\end{document}


\title{Supplementary Information for \protect\\ Topology, Vorticity and Limit Cycle in a Stabilized Kuramoto-Sivashinsky Equation}
\author{Yong-Cong Chen}
\email{chenyongcong@shu.edu.cn}
\affiliation{Shanghai Center for Quantitative Life Sciences \& Physics Department, Shanghai University, Shanghai 200444, China}
\author{Chunxiao Shi}
\affiliation{Shanghai Center for Quantitative Life Sciences \& Physics Department, Shanghai University, Shanghai 200444, China}
\author{J. M. Kosterlitz}
\email[Corresponding author: ]{j\_kosterlitz@brown.edu}
\affiliation{Shanghai Center for Quantitative Life Sciences \& Physics Department, Shanghai University, Shanghai 200444, China}
\affiliation{Permanent Address: Department of Physics, Brown University, Providence, Rhode Island 02912, USA}
\author{Xiaomei Zhu}
\affiliation{Shanghai Center for Quantitative Life Sciences \& Physics Department, Shanghai University, Shanghai 200444, China}
\author{Ping Ao}
\affiliation{Shanghai Center for Quantitative Life Sciences \& Physics Department, Shanghai University, Shanghai 200444, China}

\begin{abstract}
This supplementary information (SI) covers additional algebraic, computational and graphical calculations used or cited in the main article \cite{mainwork}.
\end{abstract}
\maketitle


\renewcommand{\vec}[1]{\mathbf{#1}}
\newcommand{\bvec}[1]{\mbox{\boldmath $#1$}}

\renewcommand{\theequation}{S\arabic{equation}}
\renewcommand{\thefigure}{S\arabic{figure}}
\renewcommand{\thesection}{\Alph{section}}
\renewcommand{\thesubsection}{\Alph{section}.\arabic{subsection}}
\newcommand{\cmmnt}[1]{}

\renewcommand{\theequation}{S\arabic{equation}}
\renewcommand{\thefigure}{S\arabic{figure}}
\renewcommand{\thetable}{S\arabic{table}}

\tableofcontents

\section{Fourier Transform}\label{sec4}

Numerical analysis of the stabilized Kuramoto-Sivashinsky (SKS) equation is most conveniently carried out in Fourier space. Following earlier work \cite{Chen23227}, we consider a one dimensional periodic lattice of length $L=Nh$ and approximate it by an {\it even} number $N$ of points with grid spacing $h$. Define the discrete forward and inverse Fourier transforms of the field $u(x)$ on a set of $N$ discrete points, $x_{n}=nh$, as
\begin{eqnarray} \label{FT1}
u_{k} & = & h\sum_{n}\,u(x_{n})\,e^{-i\frac{2\pi k\,n}{N}} = \text{fft}[u(x_{n})h](k)\\
\label{FT2}
u(x_{n}) & = & \frac{1}{L}\sum_{k}\,u_{k}\,e^{i\frac{2\pi k\,n}{N}}\;\;\;\;\;\; = \text{ifft}[u_{k}/h](n).
\end{eqnarray}
The operations $\text{fft}[\dots]$ and its inverse $\text{ifft}[\dots]$ are identical to the fast Fourier transform routines in MatLab$^{\copyright}$ with indices
\begin{eqnarray*}
n &=& 0,1, \dots, N-1 \\
k &=& 0,1,\dots,N/2-1;-N/2,\dots,-1
\end{eqnarray*}
and periodic boundary conditions $x_{N} \equiv x_{0}$ are imposed. Differential operators in $x$ space are algebraic in $k$ space so that the first derivative with respect to $x$ maps to 
\begin{equation}
\partial_{x} \rightarrow d_k = (e^{i\kappa h} -e^{-i\kappa h})/2h = d^{\ast}_{-k} = -d_{k}^{\ast}
\end{equation}
where $\kappa=2\pi k/N$ is the wavenumber of the $k^{{\rm th}}$ Fourier component. The linear operator $\hat{L}(x)$ of Eq.~\cmmnt{(\ref{linear0})}(2) maps to
\begin{multline}\label{L_k}
\hat{L}(x) \rightarrow L_{k} = \alpha + (e^{i\kappa h} -2 + e^{-i\kappa h})/h^{2}
 + (e^{2i\kappa h} - 4e^{i\kappa h} +6 -4e^{-i\kappa h} + e^{-2i\kappa h})/h^4.
\end{multline}
Note that equation numbers without the prefix ``S'' in the SI refer to the main paper \cite{mainwork}.

We define the Fourier transform of a square matrix $M(x_{n},x_{n'})$ by discrete forward and inverse transforms in $k$ space as
\begin{eqnarray}\label{FTmatrik}
M_{k,-k'} &=& h^{2}\sum_{n,n'}\,M(x_{n}, x_{n'})\,e^{-i\frac{2\pi(kn - k'n')}{N}} \\
\label{FTmatrikinv}
M(x_{n}, x_{n'}) &=& \frac{1}{L^{2}}\sum_{k,k'}M_{k,k'}\,e^{i\frac{2\pi(kn - k'n')}{N}}.
\end{eqnarray}
Note that, in the transform, the column index $k'$ has opposite sign to that of the row index $k$ in the exponents. In terms of the two-dimensional fast Fourier transform MatLab$^{\copyright}$ routines fft2[. . . ] and its inverse ifft2[. . . ], Eqs.~(\ref{FTmatrik}) and (\ref{FTmatrikinv}) are
\begin{eqnarray}\label{FTmatrik1}
M_{k,-k'} &=& \text{fft2}[M(x_{n}, x_{n'})\,h^{2}], \\
\label{FTmatrix}
M(x_{n}, x_{n'}) &=& \text{ifft2}[M_{k,-k'}/\,h^{2}].
\end{eqnarray}

In discrete $x$ space, the product of two matrices is weighted by $h$ and the unit matrix $\vec{I}$ has elements $I_{n,n'}=\delta_{n,n'}/h$. In discrete $k$ space the product of two matrices is weighted by $1/L$ so that
\begin{equation}\label{weight}
\begin{array}{ccc}
\vec{K} = \vec{M}\,\vec{N} & \;\Rightarrow \; & K_{kk'} =\frac{1}{L} \sum_{k''}M_{kk''}N_{k''k'}
\end{array}
\end{equation}
and the unit matrix $\vec{I}$ becomes $I_{kk'} = L\delta_{kk'}$. Note that matrix elements in Fourier space are complex numbers so that the Fourier transform of a real symmetric matrix in $x$ space is a Hermitian matrix in $k$ space. The transpose operation is followed by complex conjugation as usual. Sometimes, it is advantageous to work in $x$ space where the eigenvectors of real symmetric and anti-symmetric matrices are real.

\section{Stationary States of the SKS equation}

To obtain stationary periodic solutions of the SKS equation in the noiseless limit, we use the semi-implicit algorithm of \cite{Saxena_Kosterlitz}. The Fourier transform $N_{k}(t)$ of the nonlinear term $[u_{x}(x, t)]^{2}$ is found by combining the inverse and forward Fourier transforms as
\begin{equation*}
 N_{k}(t) = \text{fft}[\,(\text{ifft}[d_{k}u_{k}(t)/h])^{2}h\,](k),
\end{equation*}
so that, at time $t+\Delta t$
\begin{equation}\label{SKSF}
u_{k}(t + \Delta t) = \frac{u_{k}(t) + \Delta t N_{k}(t)}{1 + \Delta t L_{k}}.
\end{equation}
In the noiseless limit, a stationary state is achieved numerically in ${\cal N}_{t}\sim 10^{6}$ iterations with a fairly small time step, $\Delta t\sim 3\times10^{-4}$ and this final state depends on the choice of initial state. If this is dominated by a single wavenumber, $k=\pm k_{0}$ or $\kappa_{0} = 2\pi k_{0}/N$, the final state consists of components $k_{n} = \pm nk_{0}$, $n=0,1,2,\dots$, where the harmonics decay rapidly for $n>1$ and has the same periodicity as the initial state. When $ \alpha < 0.25$, there is a band of stationary states clustered around the critical or linearly fastest growing wavenumber $\kappa_{c} = 1/\sqrt{2}$ which corresponds to $\min (L_{k})$.

\section{Constructing the Global Potential}

To avoid confusion, we drop subscripts indicating the order in powers of $\tilde{u}$ when in Fourier space.The Fourier transform of $\vec{A}_{0}$ of Eq.~\cmmnt{(\ref{transfer_a})}(17) is
\begin{eqnarray}\label{A_k}
\begin{array}{cc}
A_{kk'}=2d_{k-k'}d_{k'}^{\ast}\,a_{k-k'}; & a_{k} = \text{fft}[a(x)h](k).
\end{array}
\end{eqnarray}
When $(k-k')\notin[-N/2,N/2-1]$, $(k-k')\rightarrow (k-k'\pm N)$ to ensure that $(k-k')$ is inside the defined range.
 Note that there is an error in Eq.~(34) of our earlier work \cite{Chen23227} where there is an extra factor $h$. The consequences of this error are addressed and corrected in this SI.

$\vec{Q}_{0}$ is found numerically by solving Eq.~\cmmnt{(\ref{DeltaQ2_a})}(19) using a standard MatLab$^{\copyright}$ routine, $X = \text{lyap}(A,Q)$ which solves the continuous Lyapunov equation \cite{Jbilou2006,Hached2018}, $AX+XA^{T}+Q=0$. When formatting the codes, special care must be taken in Fourier space with the $1/L$ weight associated with the product of two matrices in Eq.~(\ref{weight}). For example, the inverse of a matrix $\vec{C}$ is calculated as $L^{2}\,\text{inv}(C)$ by the matrix inversion routine of MatLab$^{\copyright}$ so that $\left[\vec{I} + \vec{Q}_{0}\right]^{-1}\Rightarrow L\,\text{inv}(I + Q_{0}/L)$. Note that the weight in $x$ space is $h$ so that $L\rightarrow 1/h$ when we transform matrices back to $x$ space. Eigenvalues and eigenmodes of $\vec{R}_{0}$ in Eq.~\cmmnt{(\ref{DeltaQPhi1_a})}(18) are computed by another standard MatLab$^{\copyright}$ routine, $[V,D] = \text{eig}(A)$. The pair-wise potential differences between the stationary states are found by using these results and Eq.~\cmmnt{(\ref{DeltaPhi4})}(22).

\subsection{Eigenmodes, Eigenvalues, and Topology}

With the {\em corrected} expression for the Fourier transform $\vec{A}_{0}$ of Eq.~(\ref{A_k}), we can correct some observations in our earlier work \cite{Chen23227}. It was stated that the number of unstable eigenmodes of $\vec{R}_{0}$ is not changed by the nonlinearity for every stationary state. Only some of these become unstable when the wavenumber $\kappa\neq \kappa_{c}$ so that the size of the unstable region agrees with that predicted by the Eckhaus instability which is a secondary instability of the periodic stationary states \cite{PhysRevE.49.166,PhysRevE.76.017204}. The remaining observations in \cite{Chen23227} remain correct. Modes with eigenvalues of small magnitude are found to lead to another stationary state inside the allowed range of $k$. This correspondence is identified by the dominant Fourier components as summarized in TABLE~\ref{Tab1} (which should replace Table~1 in \cite{Chen23227}). Note that a factor $1/L$ must be applied to the list as in Eq.~(\ref{weight}) when comparing these values to those of $L_{k}$ of Eq.~(\ref{L_k}). Hence the low-lying eigenmodes with small negative eigenvalues are seen to connect stationary states so that the topology of the web of inter-connected fixed points remains robust. This phenomenon leads to useful information about the global potential landscape as discussed below.

\begin{table}[!h]
\centering
\begin{tabular}{|c|c|c|c|c|c|c|c|}
 \hline
 \begin{tabular}{c} $k$-index \\of largest\end{tabular} &
 \multicolumn{7}{c|}{\begin{tabular}{ll}top row: & stationary state by wavenumber\\columns: & unstable eigenvalues below the state\end{tabular}} \\
  \cline{2-8}
 amplitude & $11$ & $12$ & $13$ & $14$ & $15$ & $16$ & $17$ \\
\hline
  11
 & 0.012 & 3.606 & 3.810 & 3.283 & 2.650 & 0.109 & -3.967 \\
\hline
  12
 & -1.664 & 0.000 & 26.524 & 29.903 & 1.981 & -0.352 & -4.584 \\
\hline
  13
 & -3.104 & 0.032 & 0.000 & 29.709 & 1.337 & -0.638 & -4.912 \\
\hline
  14
 & -3.583 & 0.307 & 0.745 & 0.000 & 0.626 & -0.657 & -4.740 \\
\hline
  15
 & -3.022 & 0.935 & 1.750 & 1.001 & 0.000 & -0.414 & -3.835 \\
\hline
  16
 & -1.400 & 1.931 & 4.736 & 2.469 & 32.090 & 0.000 & -2.097 \\
\hline
  17
 & 0.924 & 3.283 & 5.982 & 4.340 & 33.498 & 32.479 & 0.001 \\
 \hline
 \end{tabular}

\protect\caption{List of low-lying eigenmodes of $\vec{R}_{0}$ labeled by index of the largest component in $k$-space (1st column) and eigenvalues in columns below stationary states listed by wavenumber in units of $2\pi/L$. Conjugate $\{-k\}$ modes with identical eigenvalues are not shown. Data for $\alpha = 0.20$, $L=128$, $h=0.125$. This corrects Table~1 in the earlier work \cite{Chen23227}.}
\label{Tab1}
\end{table}

\subsection{Least Squares Polynomial Fitting}

With this novel topology in mind, we first evaluate the pairwise potential difference between two stationary states from Eq.~\cmmnt{(\ref{DeltaPhi4})}(22). 
As a first attempt, we use the potential differences between pairs of neighboring stationary states to map out incrementally the potential $\Phi(\kappa)$ over the whole range of $\kappa$, assuming that this potential exists. However, this leads to a monotonically increasing potential as the wavenumber $\kappa$ increases which implies that we must consider the topology of the set of all states. Note that Fig. 3 in \cite{Chen23227} is not accurate and is no longer meaningful.

\begin{figure}[!ht]
\begin{tabular}{ccc}
\includegraphics[width=0.3\textwidth]{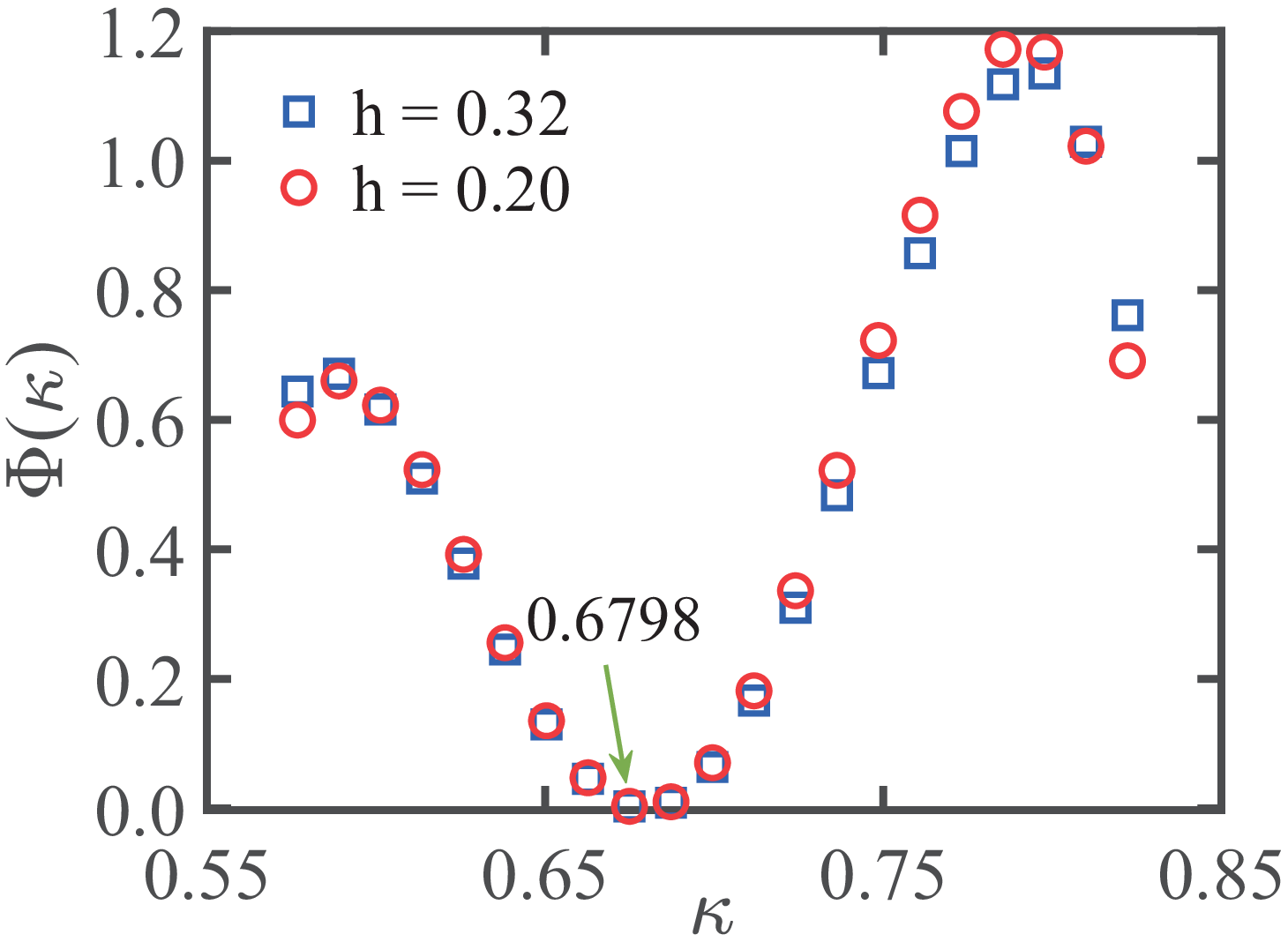} &
\includegraphics[width=0.3\textwidth]{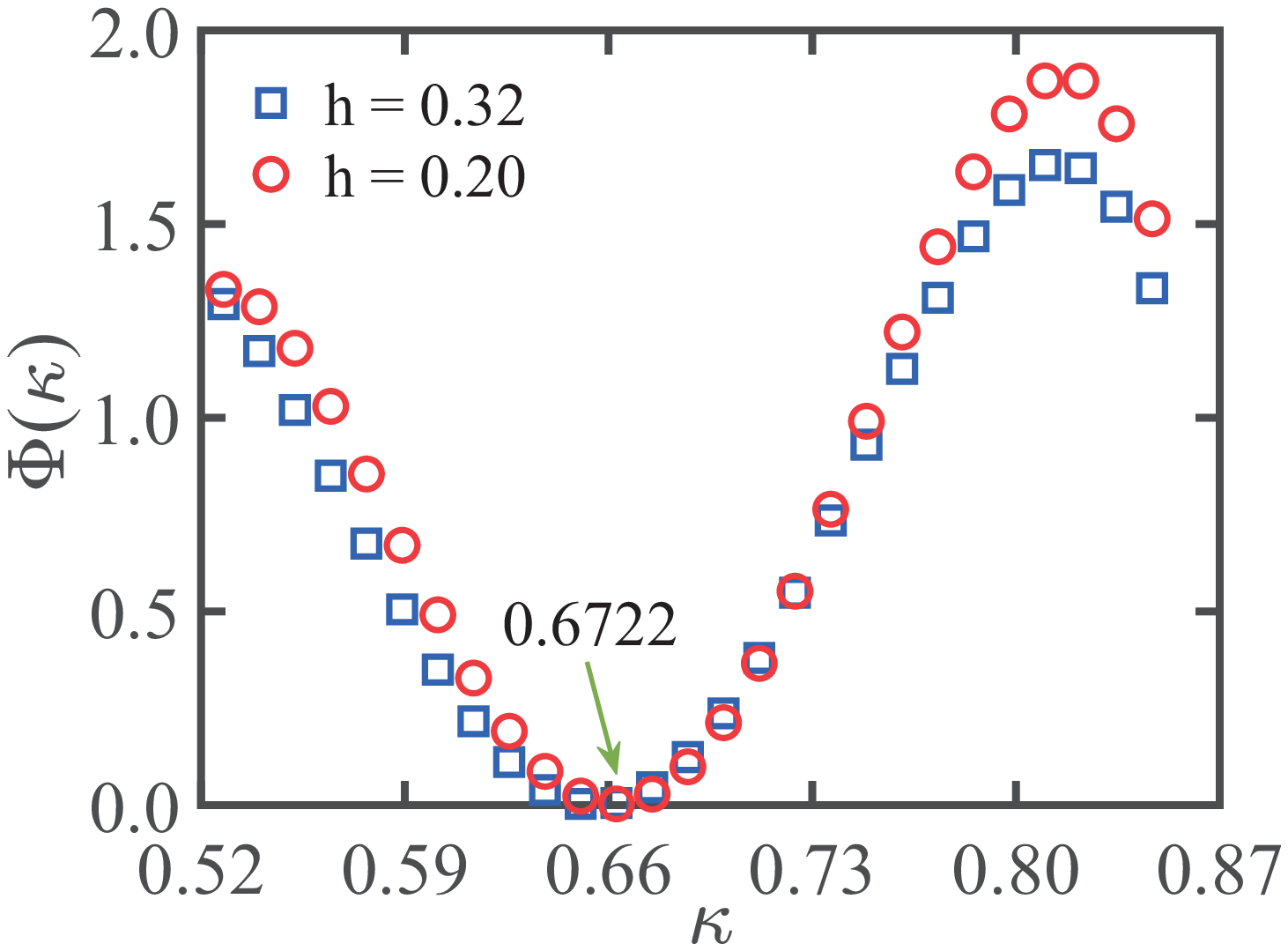} &
\includegraphics[width=0.3\textwidth]{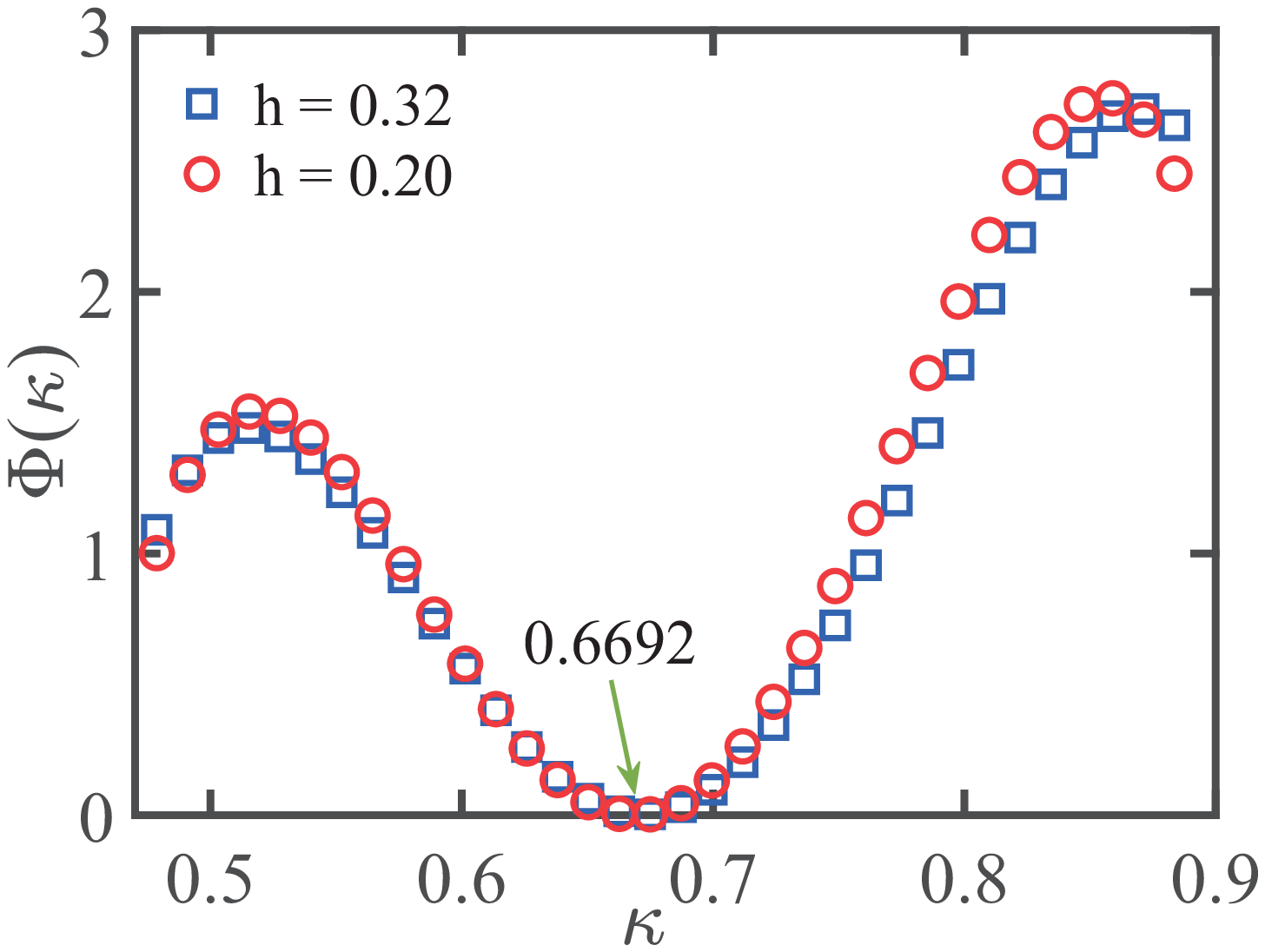} \\
(a) $ \alpha = 0.22 $ & (b) $ \alpha = 0.20$ & (c) $ \alpha = 0.17$
\end{tabular}

\protect\caption{Global potentials vs. wavenumber ($\kappa$) with $L = 512$ from a $4^{\rm th}$ order polynomial fitting over the entire topology of the stationary states for several values of control parameter $\alpha$. Arrows mark the most probable wavenumber for each $\alpha$.}\label{SFig1}
\end{figure}

There are many paths in state space connecting any chosen pair of fixed points and each path involves different intermediate states. Every path yields a different potential difference between the same two states from the pair-wise potential formula so that this procedure does not yield a potential but rather an action. The {\it true} potential difference must be path independent according to Eqs.~\cmmnt{(\ref{DeltaQ})}(12) and \cmmnt{(\ref{DeltaPhi1})}(13) and one way to overcome this problem of path dependence is to weight the paths by a fitting procedure. We assume the existence of a potential which is parameterized by a low-order polynomial in the wavenumber $\kappa$ and optimize the coefficients over all pair-wise potential differences. The resulting potential is insensitive to the order of the polynomial for $4^{\rm th}$, $5^{\rm th}$ and $6^{\rm th}$ order fits since these are almost identical, thus validating the fitting procedure. This new result, shown in Fig.~\ref{SFig1}, replaces Fig.~4 of \cite{Chen23227}. Note that the number of nearest neighbors used in \cite{Chen23227} as a fitting parameter is not needed here.

\section{Stochastic Simulations}

\subsection{Adding Noise to the SKS Equation}

The global potential landscape can be checked by direct, brute-force stochastic simulations. Of necessity, these simulations must be done with rather strong noise so that {\it all} stationary states have some finite probability of being visited during the simulation. With noise of strength $\epsilon$, the evolution Eq.~(\ref{SKSF}) is modified to
\begin{equation}\label{SKSF1}
u_{k}(t + \Delta t) = \frac{u_{k}(t) + \Delta t N_{k}(t) + \sqrt{2\epsilon \Delta t N\,h}\,\xi_{k}(t)}{1 + \Delta t L_{k}}.
\end{equation}
where $\xi_{k}(t)$ is the Fourier transform of a Gaussian distributed random variable with zero mean, $\langle\xi(t)\rangle=0$, and uncorrelated in time and space so that
\begin{equation}\label{SKSF2}
\langle \xi_{k}(t)\xi^{\ast}_{k'}(t')\rangle = \delta_{k,k'}\delta_{t,t'}.
\end{equation}
Technically, for every $k\neq 0$, we use two sets of real normal noises $\xi^{(1)}_{k}, \xi^{(2)}_{k}$ so that $\xi_{k} = [\xi^{(1)}_{k}+i\,\text{sgn}(k)\,\xi^{(2)}_{k}]/\sqrt{2}$. Note that our Fourier transform convention of Eq.~(\ref{FT1}) has an overall factor $h$, so that the same factor appears in the numerator of Eq.~(\ref{SKSF1}), rather than in the denominator of the noise term (cf. \cite{Saxena_Kosterlitz}).

\subsection{Probability Distribution of Occupancies}

In a simulation with noise, the probability of the system being in a given stationary state is a difficult question because, by definition, a stationary state is a solution of the {\it noiseless} SKS equation and, in a simulation without noise, the final stationary state is determined by the initial state of the system. To address this issue, we take a set of snap shots of the system evolving with external noise at periodic time intervals. To evolve these states to their associated stationary states, we use one of the snapshots as an initial state of the system which then evolves {\it deterministically} for time $\Delta T_{s}$ to some stationary state which depends on the choice of $\Delta T_{s}$. By performing this for every snapshot, we can define the probability distribution $P_{s}(\kappa)\propto \exp[-\Phi_{s}(\kappa)/\epsilon]$ (but cf. below) of finding the system in the stationary state $\kappa = 2\pi |k|/L)$. This scheme is used to verify the theoretical predictions from the global potential landscape point of view.

The most stable state is expected to correspond to a minimum of the potential. The global potential $\Phi(\kappa)$ obtained from the stochastic decomposition is compared to the $\Phi_{s}(\kappa)$ from simulations. The $P(\kappa)$ should be a standard Boltzmann distribution \cite{Ao_Thouless} so one expects $P(\kappa) = P_{0}(\kappa)\exp[-\Phi(\kappa)/\epsilon]$ where $\epsilon$ is the noise strength of Eq.~\cmmnt{(\ref{diffusion})}(3) and $P_{0}(\kappa)$ varies slowly with $\kappa$. We now compare $\Phi_{s}(\kappa)$ with $\Phi(\kappa)$.

\begin{figure}[!ht]
\begin{tabular}{cc}
\includegraphics[width=0.245\textwidth]{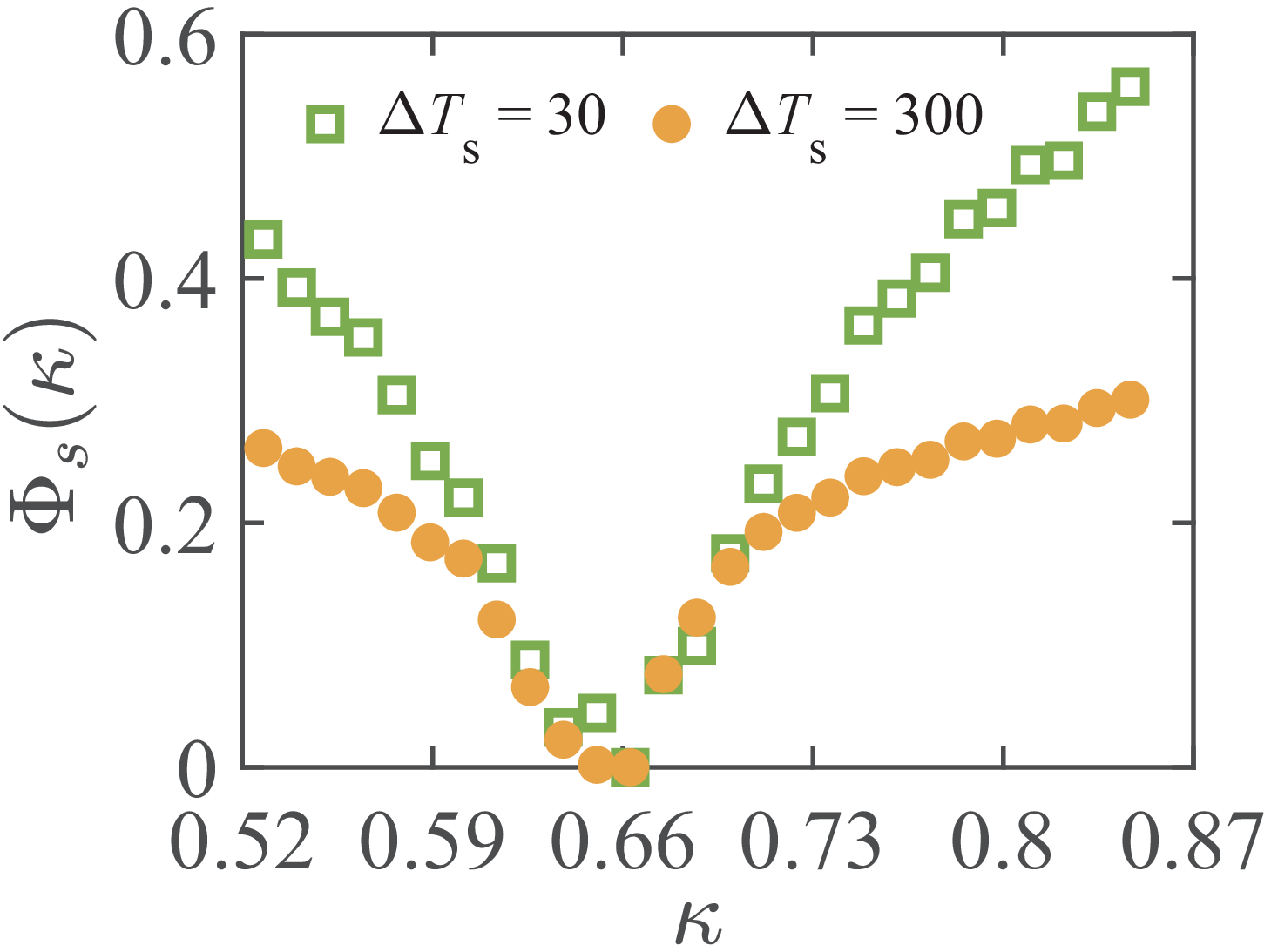}
\includegraphics[width=0.245\textwidth]{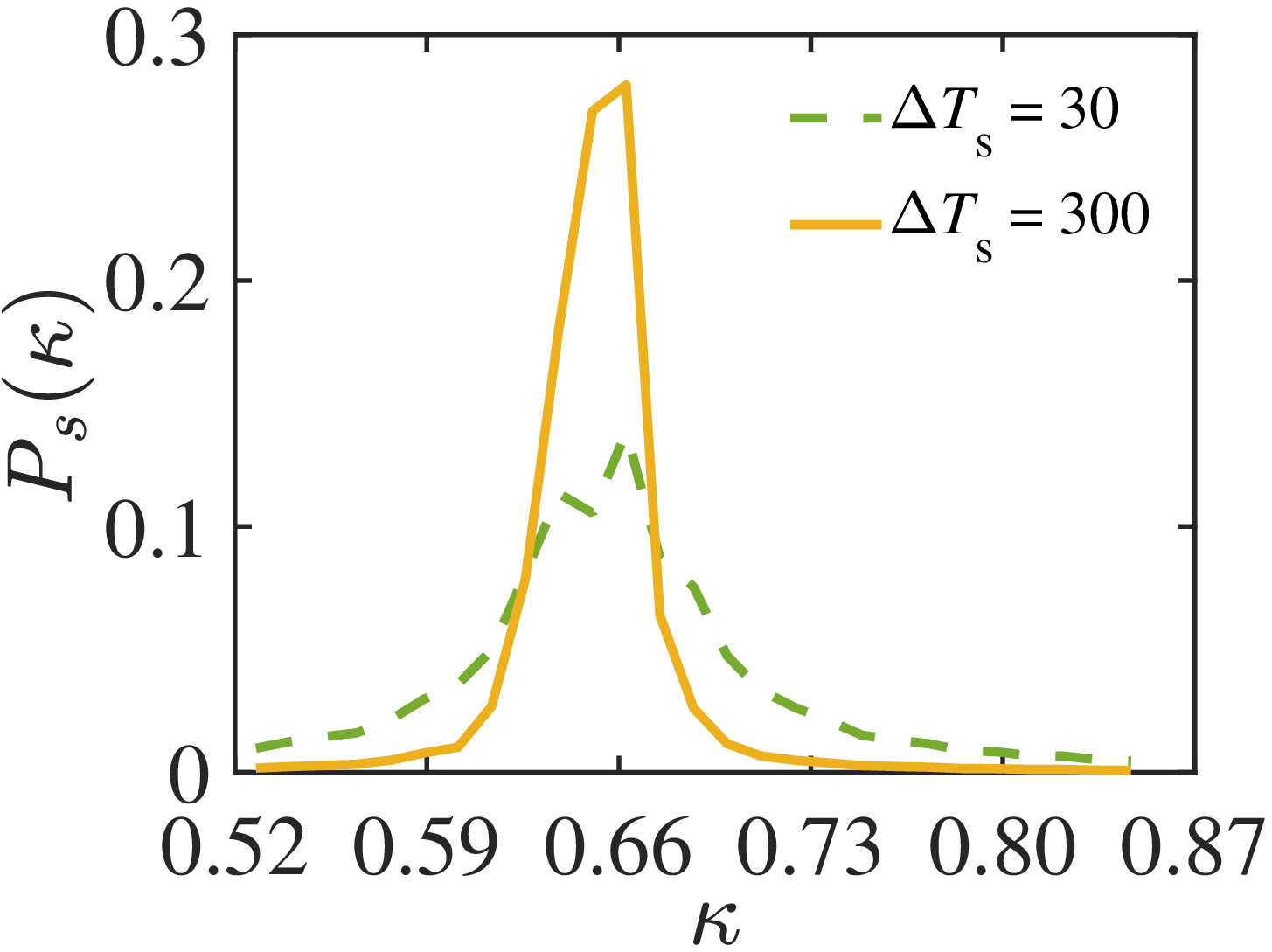} &
\includegraphics[width=0.245\textwidth]{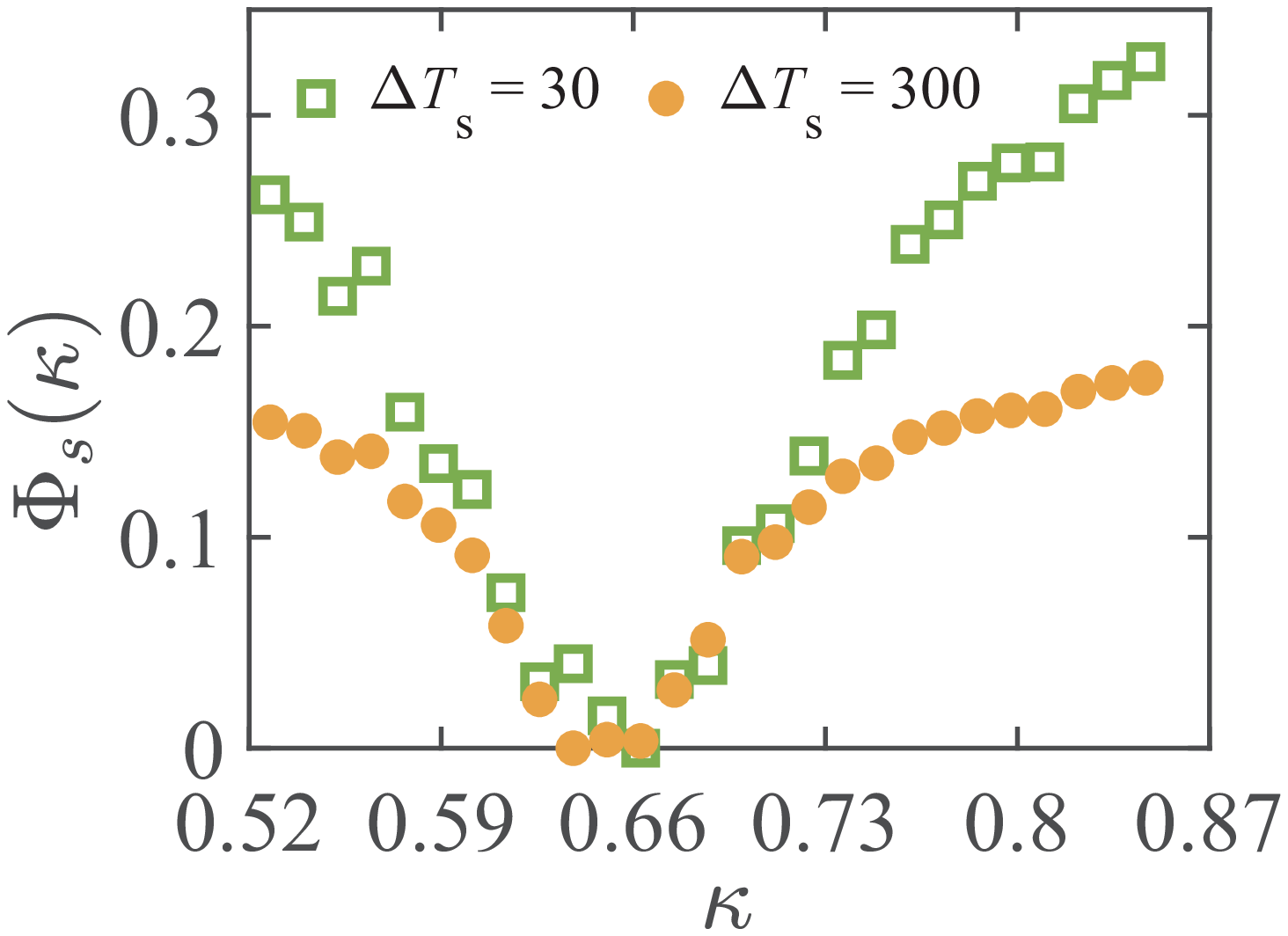}
\includegraphics[width=0.245\textwidth]{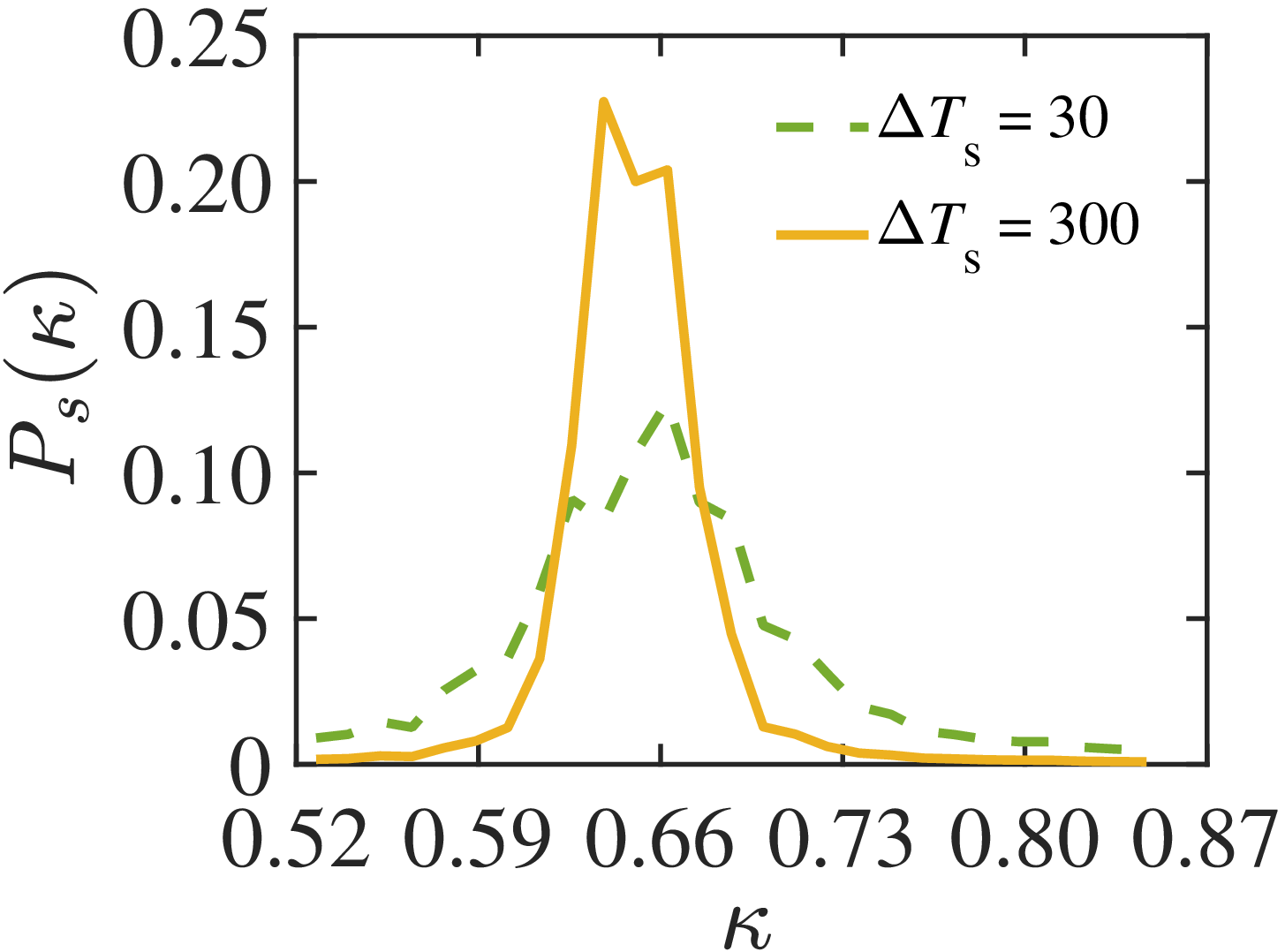} \\
(a) $\epsilon = 0.8$ & (b) $\epsilon = 0.3$  \\
\includegraphics[width=0.245\textwidth]{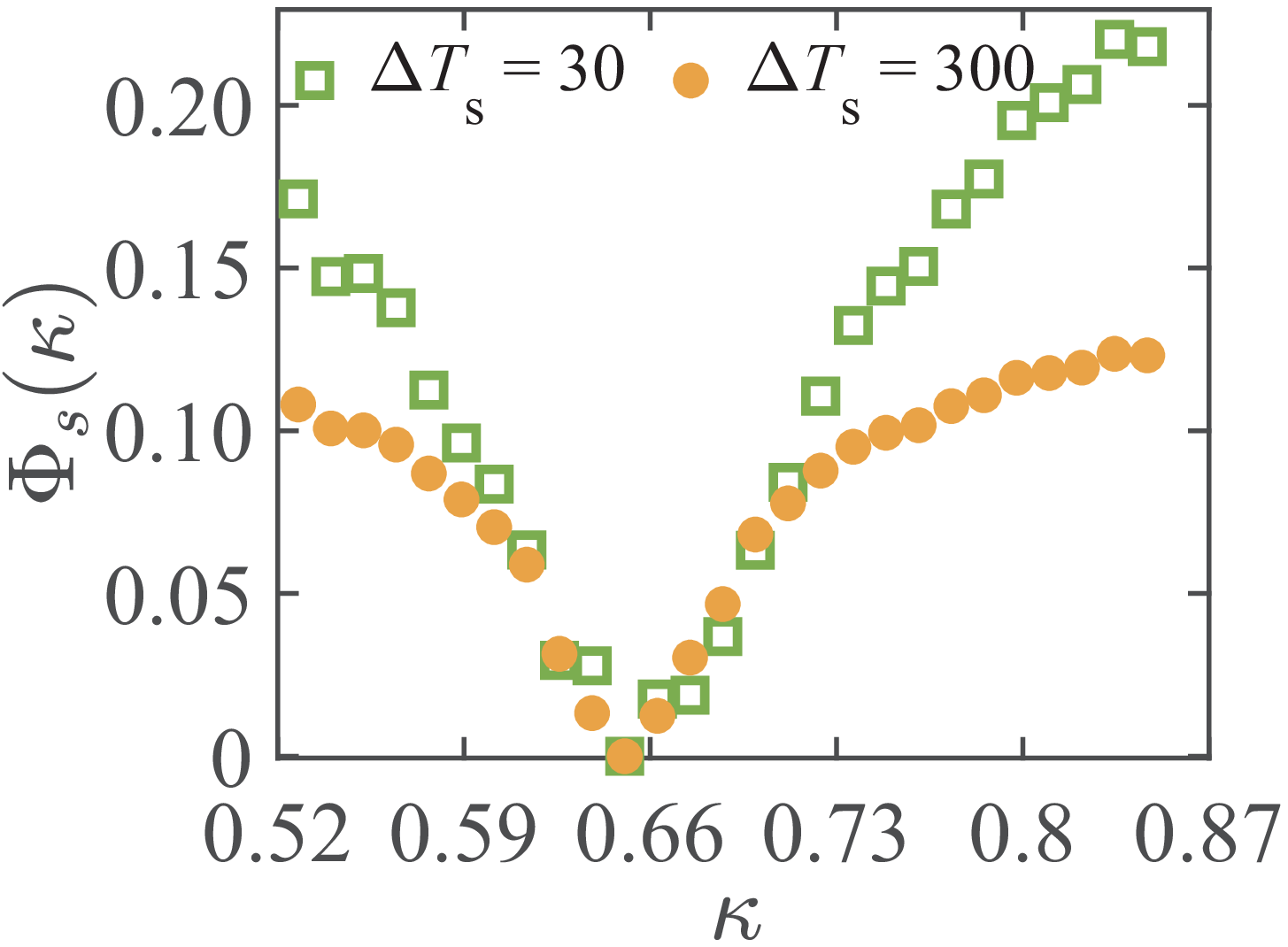}
\includegraphics[width=0.245\textwidth]{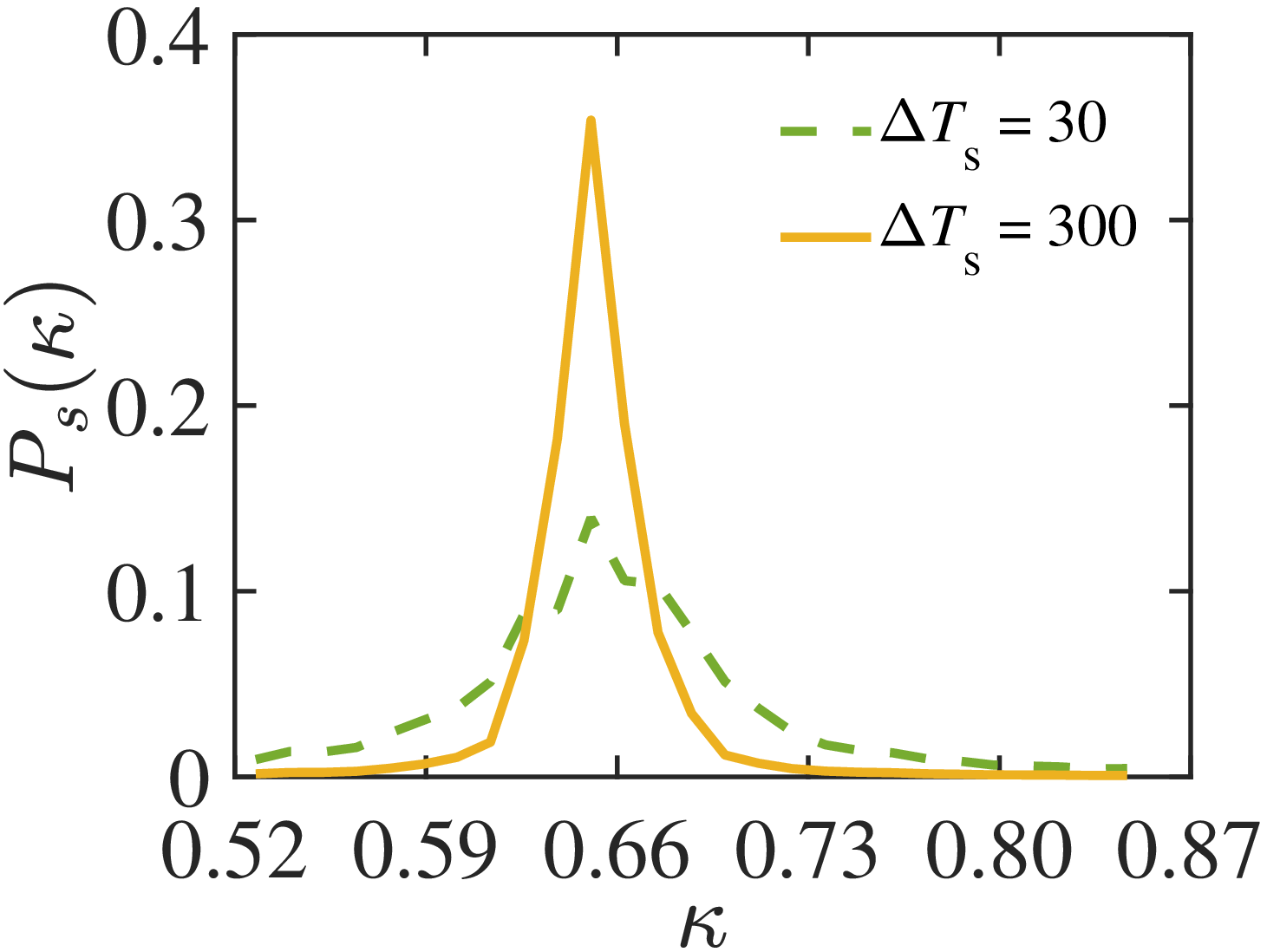} &
\includegraphics[width=0.245\textwidth]{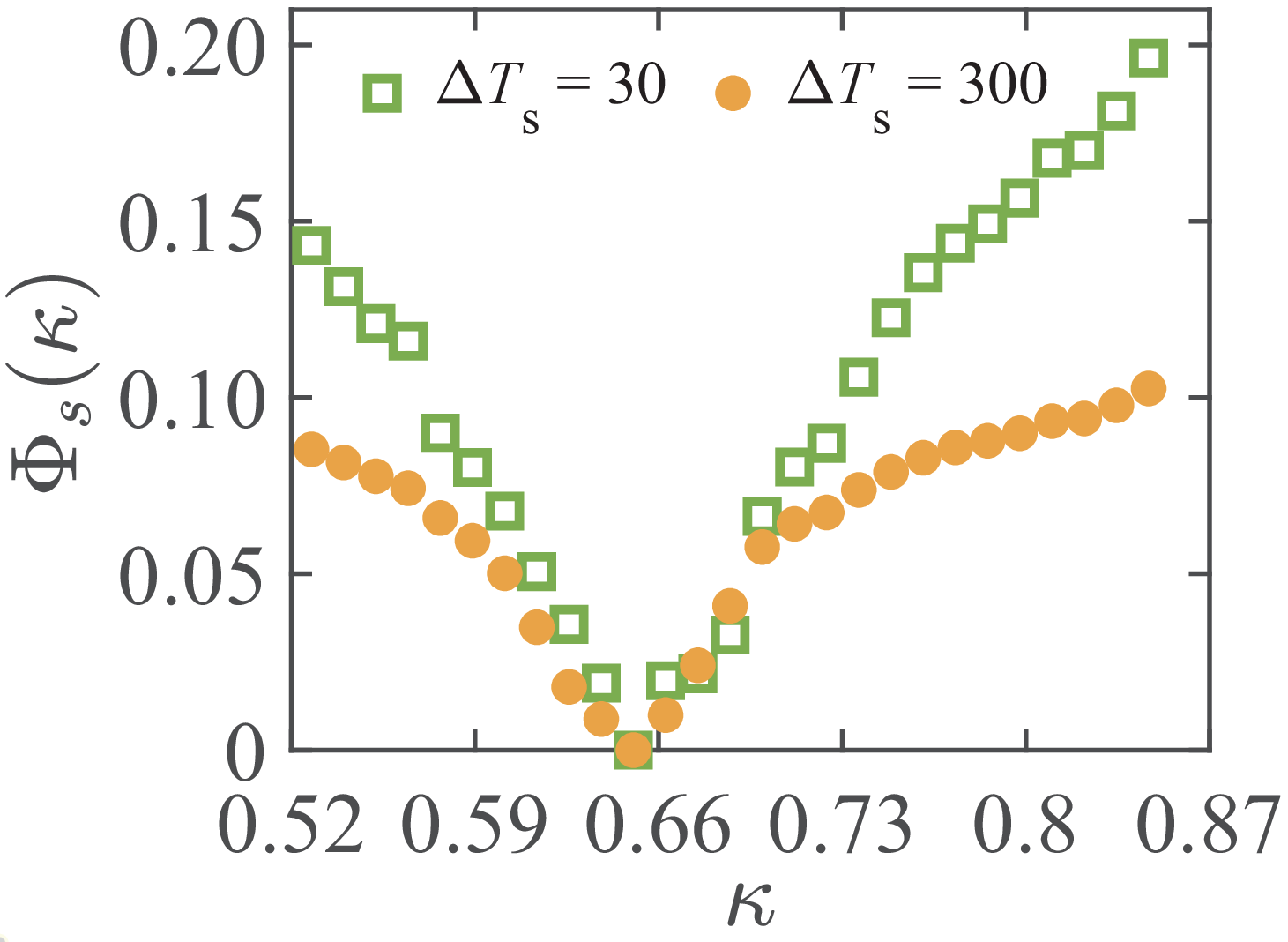}
\includegraphics[width=0.245\textwidth]{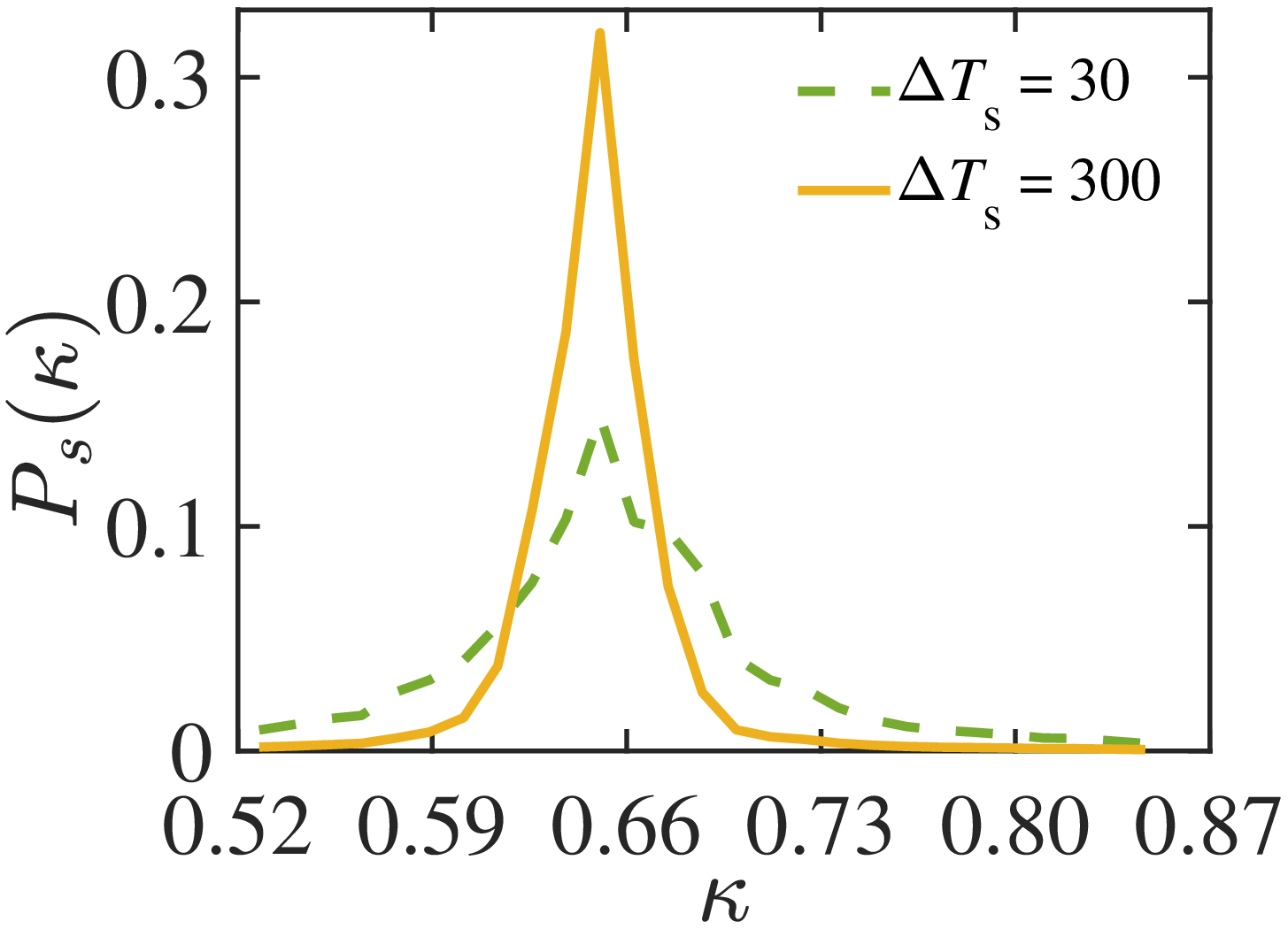} \\
(c) $ \epsilon = 0.12 $ & (d) $ \epsilon = 0.08$
\end{tabular}
\protect\caption{Simulated potentials $\Phi_{s}(\kappa)$ (left), obtained from the probability distribution $P_s(\kappa)$ with $\tilde{\epsilon} = \sqrt{\epsilon/\Delta T_{s}}$ (right), for $L = 512$, $h = 0.32$, $\alpha = 0.20$, and different noise strengths $\epsilon$.}\label{SFig2}
\end{figure}

It turns out that $\Phi_{s}(\kappa)$, shown in Fig.~\ref{SFig2}, does agree with the position of the minimum and also with the shape of $\Phi(\kappa)$ when the noise strength $\epsilon$ is replaced by a rescaled effective noise strength $\tilde{\epsilon} = \sqrt{\epsilon/\Delta T_{s}}$. This $\Phi(\kappa)$ correctly predicts the most stable state and the shape of the potential up to some overall scale factor. This holds independently of the individual values of $\epsilon$ and the relaxation time $\Delta T_{s}$ which supports the theoretical analysis although neither theoretical nor numerical analysis is able to predict the absolute scale of the global potential landscape. This discrepancy remains to be investigated.

\section{Vortex Oscillation and Limit Cycle}

Vortex like oscillations and limit cycle motion of the eigenstates in $\kappa$ space are predicted from general considerations based on the stochastic dynamics as discussed in the main work \cite{mainwork} and are observed near stationary points. The matrix $\vec{Q}$ is found from Eq.~\cmmnt{(\ref{DeltaQ2_a})}(19) in Fourier space (we drop the subscript 0 for simplicity) and, to obtain real matrices, we work in coordinate space by using the transform of Eq.~(\ref{FTmatrix}). The eigenmodes and eigenvalues of $\vec{Q}$ are computed by a standard routine in MatLab$^{\copyright}$, $[V,E] = \text{eig}(A)$, which gives pairs of pure imaginary eigenvalues of the matrix $E$. To obtain real eigenvalues in the form $q_{i}\,(i\bvec{\sigma}_{y})$, which are real $2\times 2$ antisymmetric matrices used in the main work \cite{mainwork}, we perform a rotation of $V$ by $Y=V*P$ where $P$ consists of $N/2$ diagonal $2\times 2$ matrix blocks of $(\vec{1}-i\,\bvec{\sigma}_{x})/\sqrt{2}$ and $\bvec{\sigma}_{x}$ where $\bvec{\sigma}_{y}$ are Pauli matrices. The eigenvectors obtained from this form the basis for Eq.~\cmmnt{(\ref{lc0})}(25), labelled by the eigenvalues $\{q_{i}\}$ in descending order as discussed in Eqs.~\cmmnt{(\ref{matrix1})}(23)- \cmmnt{(\ref{lc0})}(25). Each of these pairs constitutes a $2D$ subspace in which the system can have circular oscillations about a stationary state.

The first few pairs in each steady state are the most likely to exhibit this vorticity. We evolve the system from an initial state with a finite amplitude of such a pair, which is in turn a small perturbation on the underlying steady state and quasi periodic oscillations or vortex motions around some fixed points are indeed observed. This behavior can occur at both stable and unstable fixed points. In the case of a fixed point which is unstable in both dimensions of the subspace, the oscillation moves away from the fixed point, leading to the possibility of a limit cycle when the nonlinearity becomes sufficiently large. An example of this is discussed in the main work \cite{mainwork} and more examples are discussed below.

\subsection{Typical Oscillations and Regions with Vorticity}

\begin{figure}[!ht]
\begin{tabular}{ccc}
\includegraphics[width=0.33\textwidth]{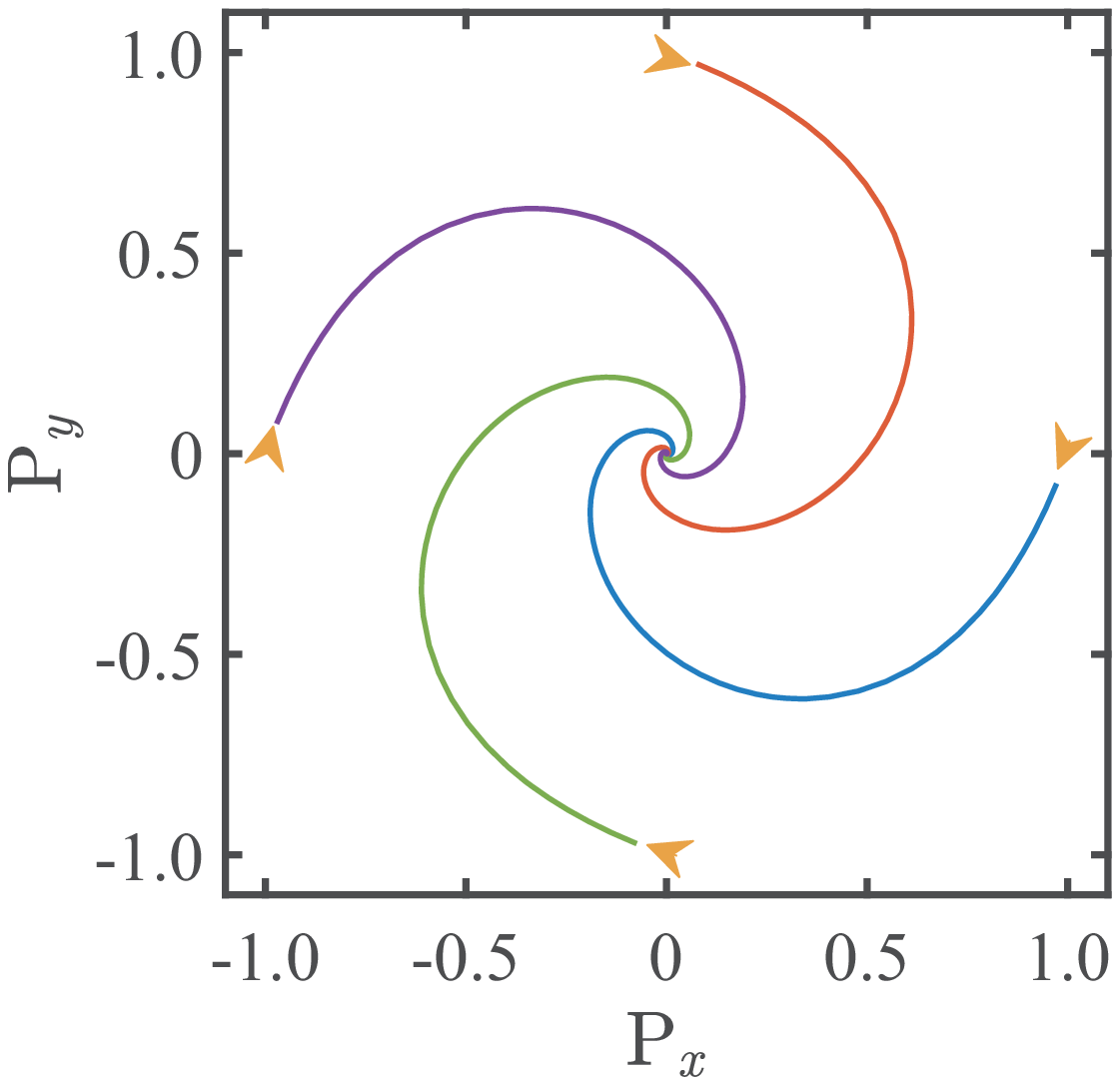} &
\includegraphics[width=0.33\textwidth]{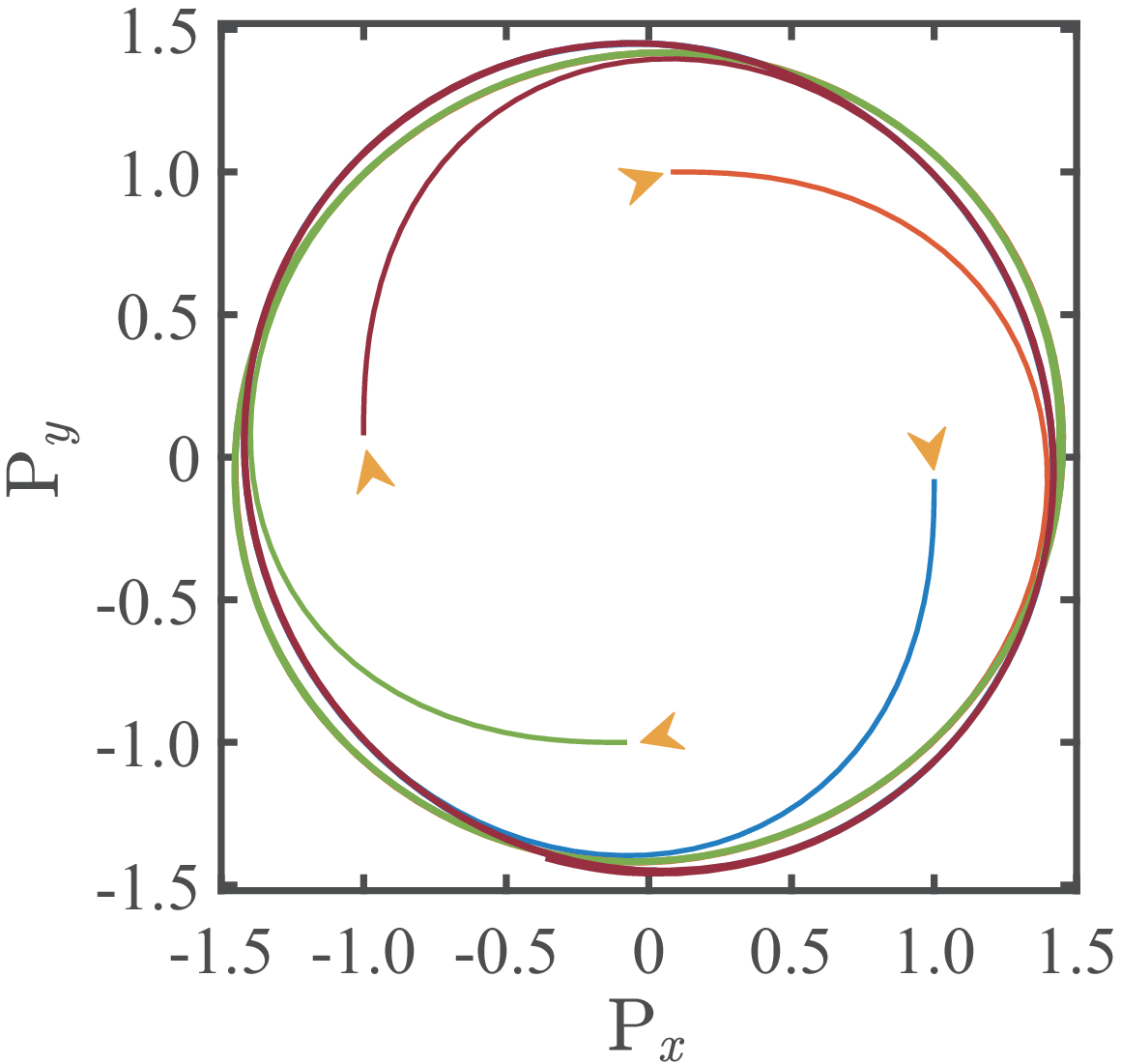} &
\includegraphics[width=0.33\textwidth]{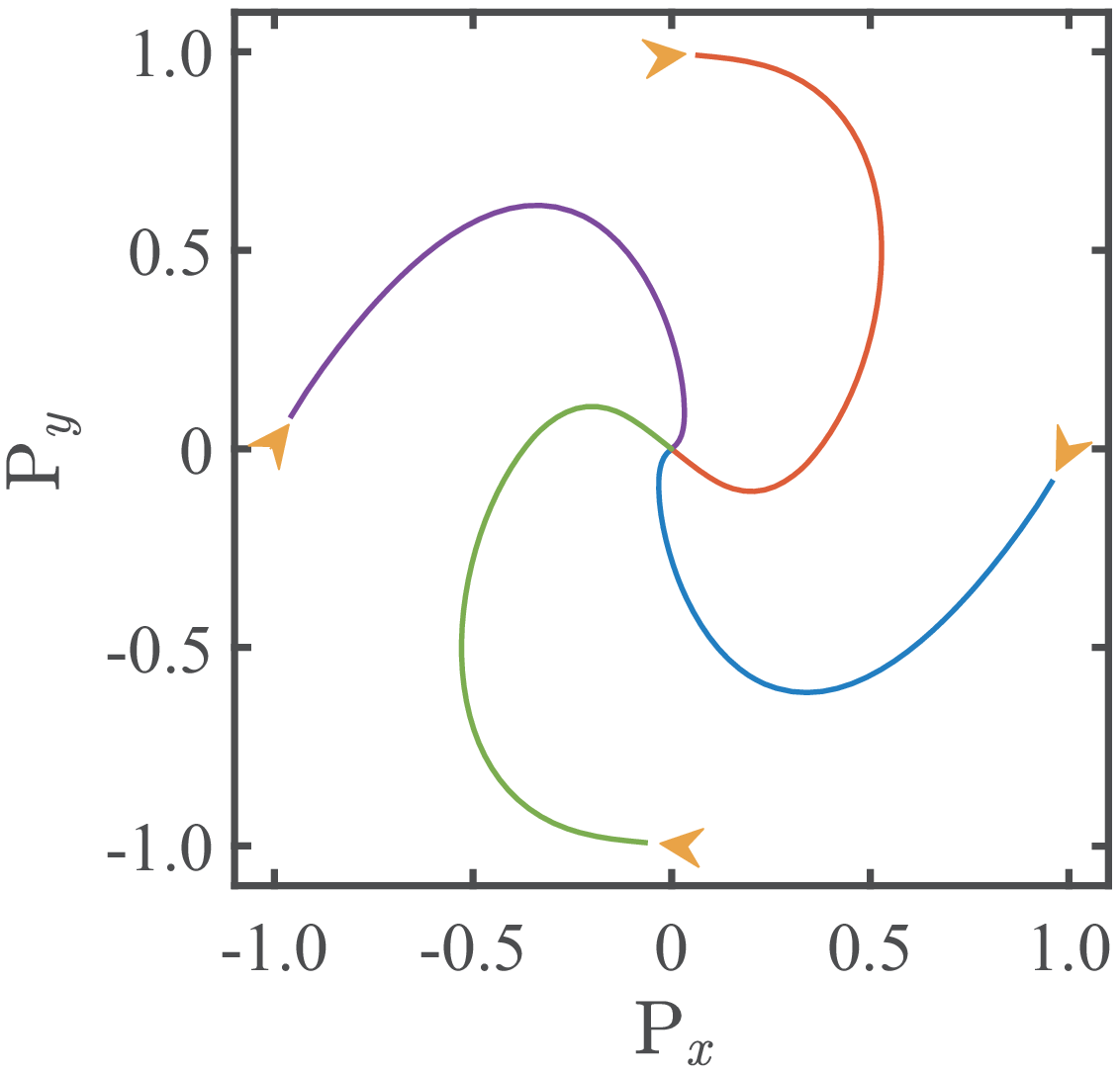} \\
(a) $\kappa = 0.6995$ & (b) $\kappa = 0.6995$ & (c) $\kappa = 0.7363$
\end{tabular}
\protect\caption{Trajectories ${\rm{P}}_{\sigma}(t) = \vec{e}^{\dag}_{i\sigma}\cdot\tilde{\vec{u}}(t)$ in 2D dimensional subspace for $L = 512$, $h = 0.32$, $\alpha = 0.20$ with different wavenumber. (a) Small deviations from a stable state decay back to the origin with vortex like circulation for $\kappa = 0.6995$. (b) Vortex like motion seen by removing attenuation from the trajectories with the same parameters as (a). (c) Small deviations from a stable state decay to zero {\it without} vortex like circulation for $\kappa = 0.7363$.}\label{SFig3}
\end{figure}

A typical vortex like motion is shown in Fig.~\ref{SFig3} (a),(b) near a stationary periodic state inside the Eckhaus stable region, with $L = 512$, $h = 0.32$, $\alpha = 0.20$ and $\kappa = 0.6995$. Motion in the $15^{\rm th}$ subspace is described by Eq.~\cmmnt{(\ref{lc0})}(25) and we impose a small initial deviation from the stationary state in the direction of the eigenvector by $\tilde{\vec{u}}_{0}=\pm\vec{e}_{15\sigma}$ ($\sigma=1, 2$) in Fourier space. Each state then evolves according to Eq.~(\ref{SKSF}) and the state vector is projected back onto the $i^{{\rm th}}$ subspace by 
{${\rm{P}}_{\sigma}(t) = \vec{e}^{\dag}_{i\sigma}\cdot\tilde{\vec{u}}(t)$ (rename $\sigma$ to $x,y$ for convenience)}. When the exponential damping factors on the trajectories are removed, quasi circular motions appear as shown in Fig.~\ref{SFig3} (b). However, at high wave numbers, shown in Fig.~\ref{SFig3} (c) for the $1^{\rm st}$ subspace, the vorticity disappears so that trajectories in this subspace converge to the stable state but {\it without} vortex like circulation.

\begin{figure}[!ht]
\includegraphics[width=0.5\textwidth]{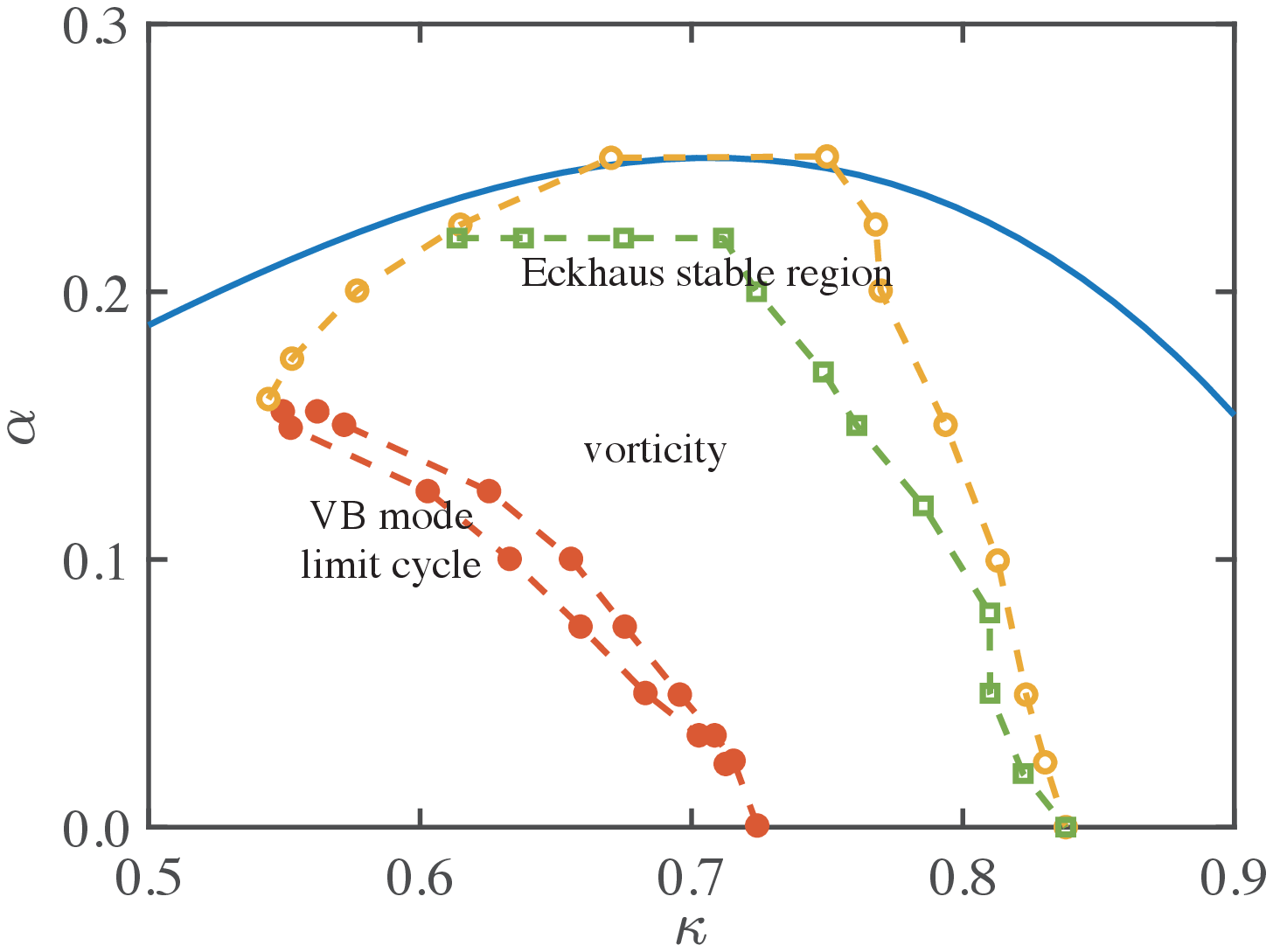}
\protect\caption{Phase diagram for some parameters. In a vacillating-breathing (VB) mode, there will be quasi periodic oscillation in the region far away from the fixed point, namely, limit cycle motion. In the range of stable region, the region below the green dotted-square (contains the points on this line) shows stable vortex like circulation, which is absent at large wavenumber.}\label{SFig4}
\end{figure}

We next examine parameter regions which exhibit vorticity. We also investigate numerically the degree of isolation of pairs of eigenmodes from Eq.~\cmmnt{(\ref{overlap})}(28) and Fig.~\ref{SFig4} gives an overall picture of this. In general, vorticity exists to the left of the dotted-square line.
It is of interest to note that this dividing line coincides with the minima of mode isolation. The degree of isolation or the pair approximation improves as $\alpha$ decreases, even for unstable states. The same happens on the right-hand side of the line, although vorticity is absent for large wavenumbers. At present, there is no analytical understanding of this.

\subsection{Vorticity Driven VB Modes and Pattern Drifting}

\begin{figure}[!ht]
\begin{tabular}{c}
\includegraphics[width=0.8\textwidth]{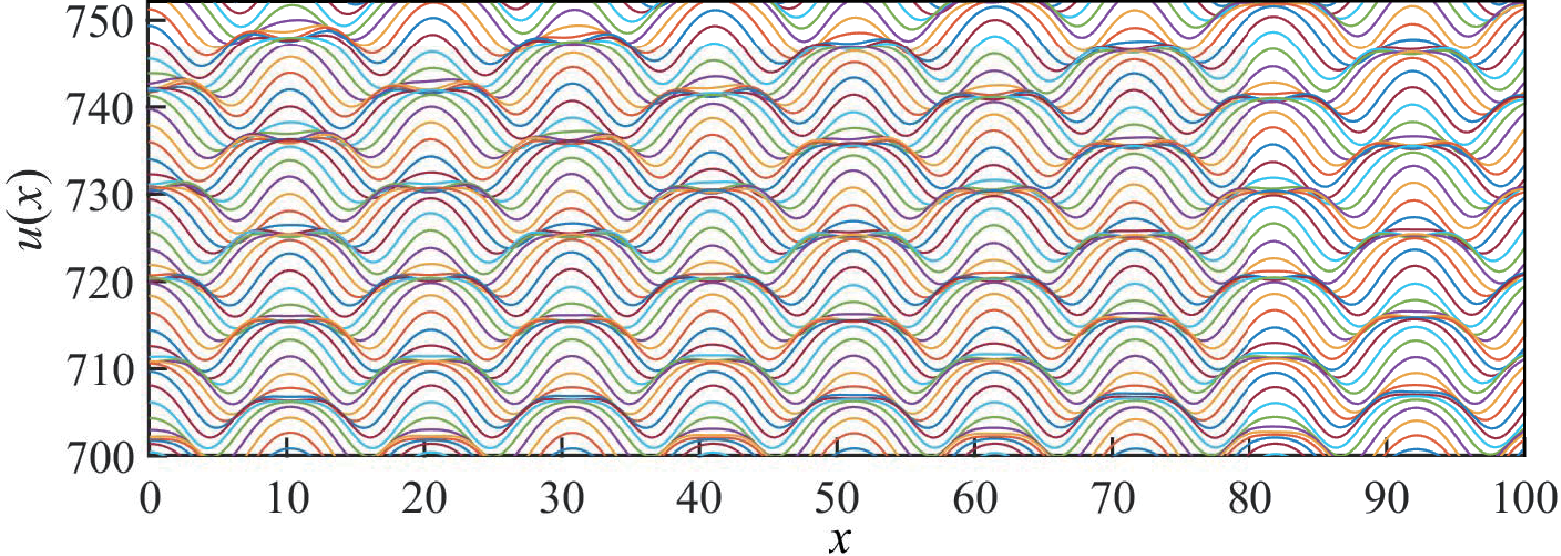} \\
(a)
\end{tabular}
\begin{tabular}{cc}
\includegraphics[width=0.4\textwidth]{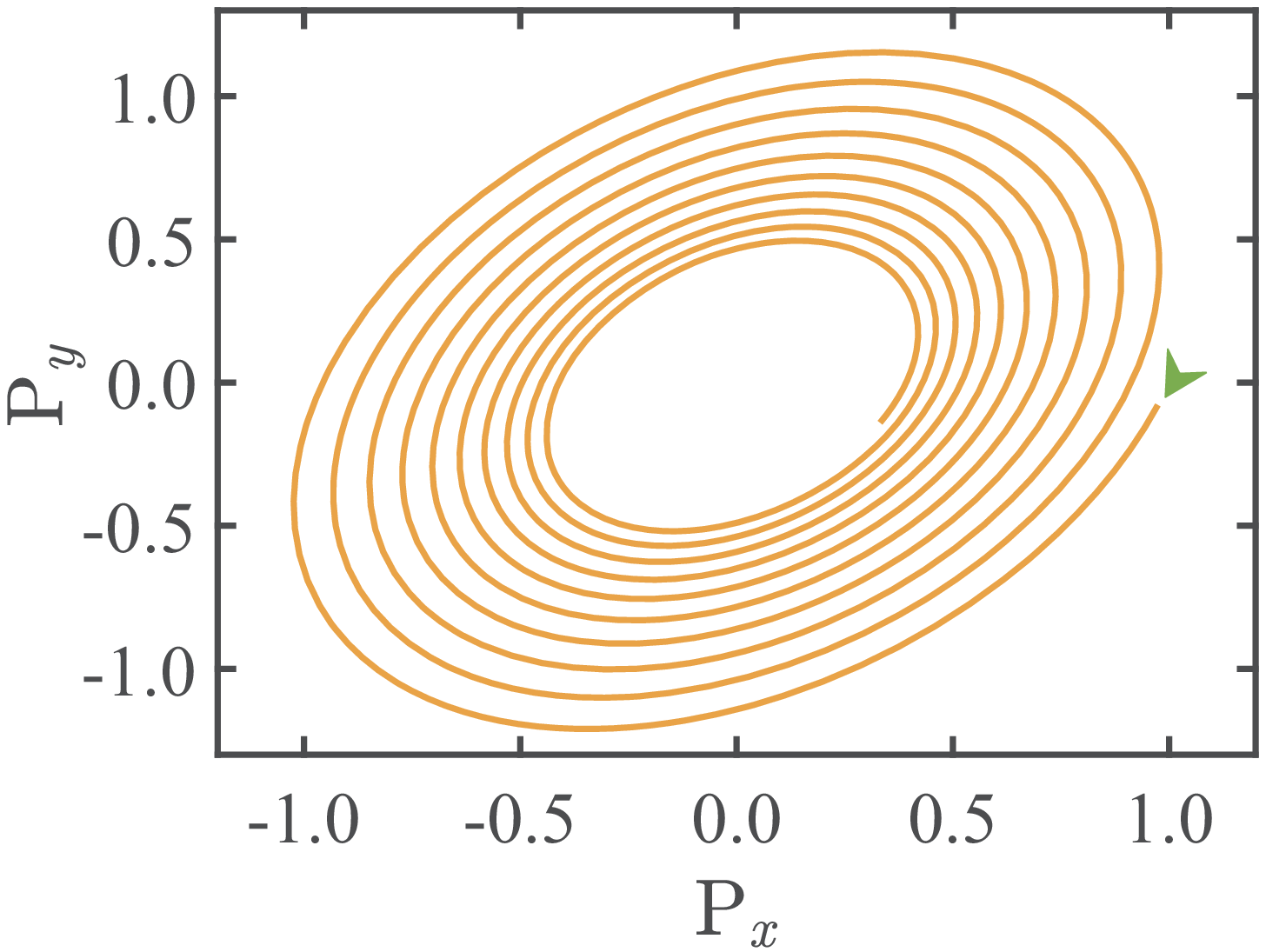} &
\includegraphics[width=0.4\textwidth]{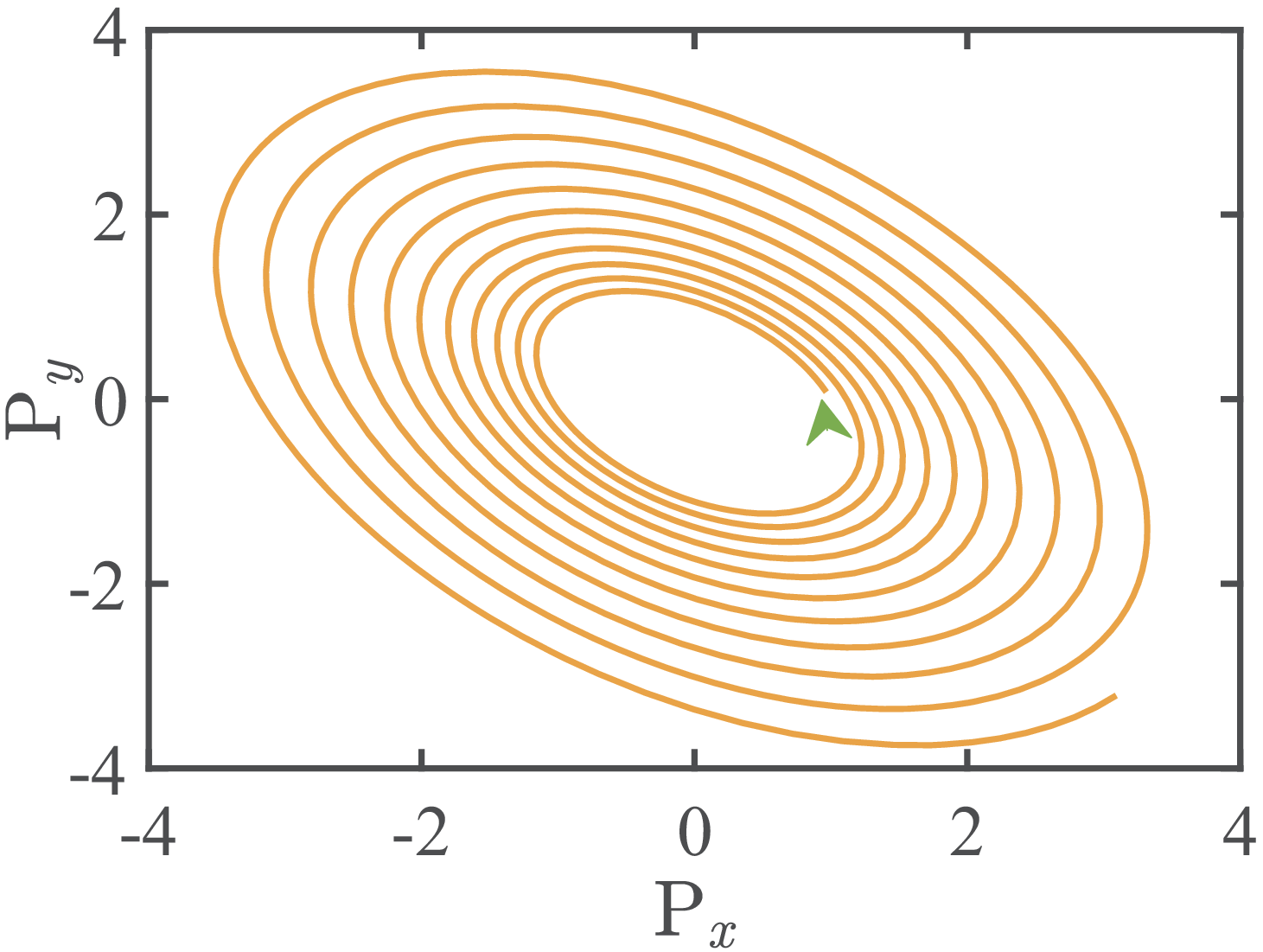} \\
(b)  & (c)
\end{tabular}
\protect\caption{Vortex motions in VB oscillations decomposed into the eigenstates of $\vec{Q}_{0}$ with $L = 512$, $h = 032$, $\alpha = 0.12$ and cellular wavenumber $\kappa = 0.6136$. (a) Spatiotemporal portrait of the oscillations, showing no apparent vorticity in the dynamics. (b) Decaying circulation in the stable $3^{\rm rd}$ subspace. (c) Growing circulation in the unstable $5^{\rm th}$ subspace. Here ${\rm{P}}_{\sigma}(t) = \vec{e}^{\dag}_{i\sigma}\cdot\tilde{\vec{u}}(t)$.}\label{SFig5}
\end{figure}

A vacillating-breathing (VB) mode \cite{Kerszberg1983,Obeid_Kosterlitz,Cross2016,Saxena_Kosterlitz} in which each cell oscillates out of phase by $\pi$ with its neighbors occurs around $\alpha\cong 0.1$ and $\kappa < \kappa_{c}$. This is the ideal region to show that the VB mode corresponds to the vortex motion discussed earlier and also to study the mechanism causing vorticity to drive the phase drifting of a periodic stationary state. {A typical scenario of VB oscillation is shown in FIG.~\ref{SFig5}.}

\begin{figure}[!ht]
\begin{tabular}{cc}
\includegraphics[width=0.4\textwidth]{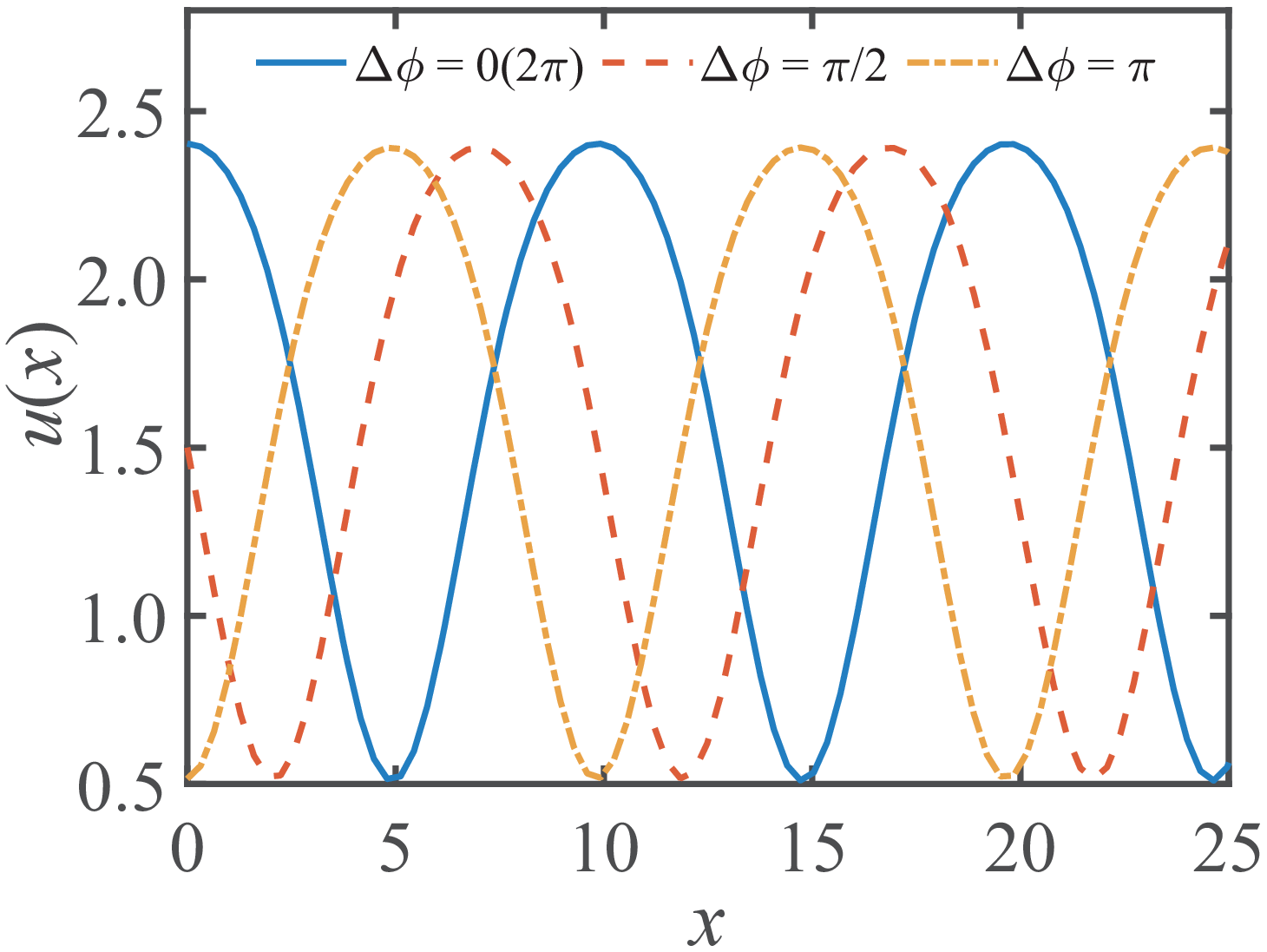} &
\includegraphics[width=0.4\textwidth]{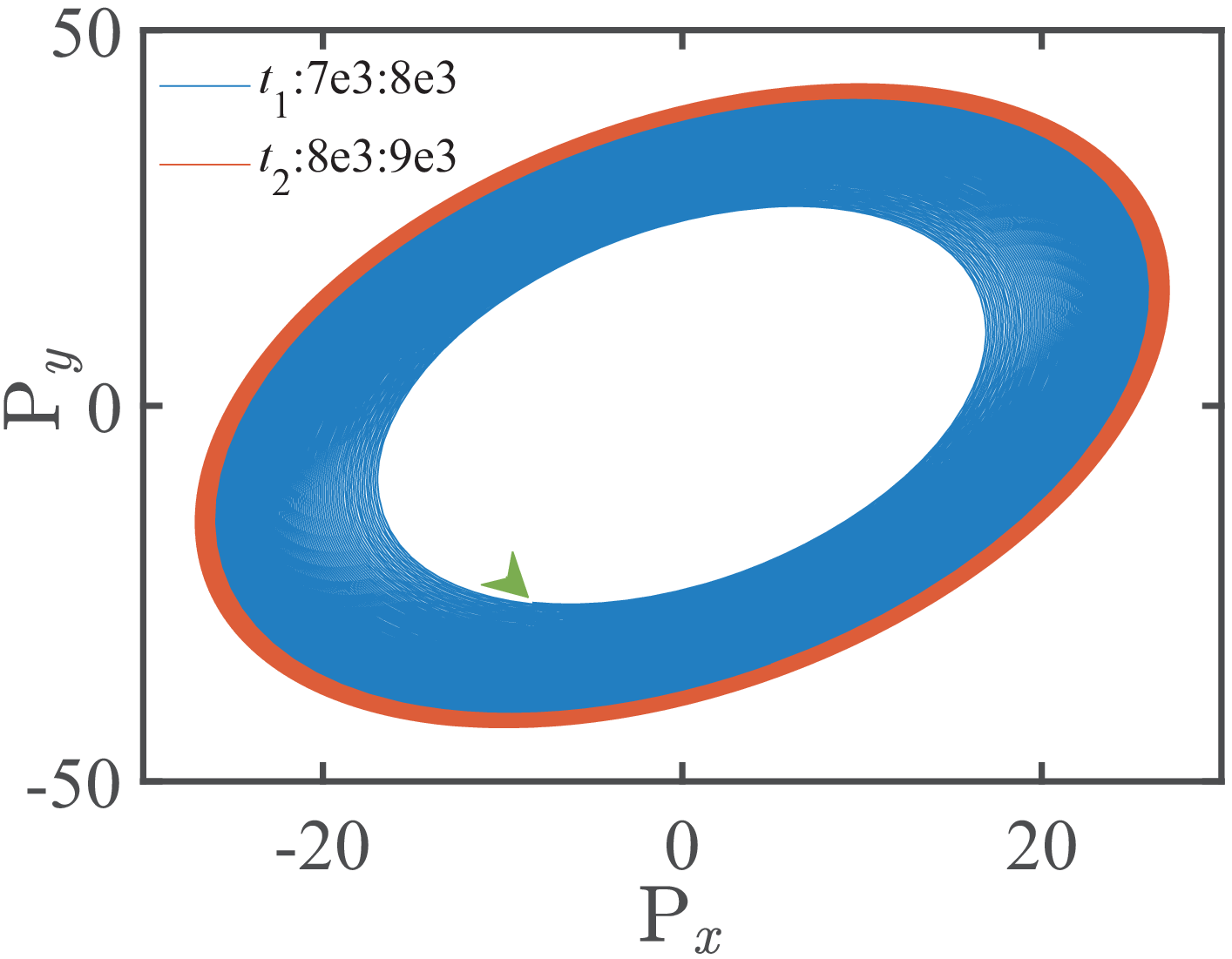} \\
(a)  & (b) \\
\includegraphics[width=0.4\textwidth]{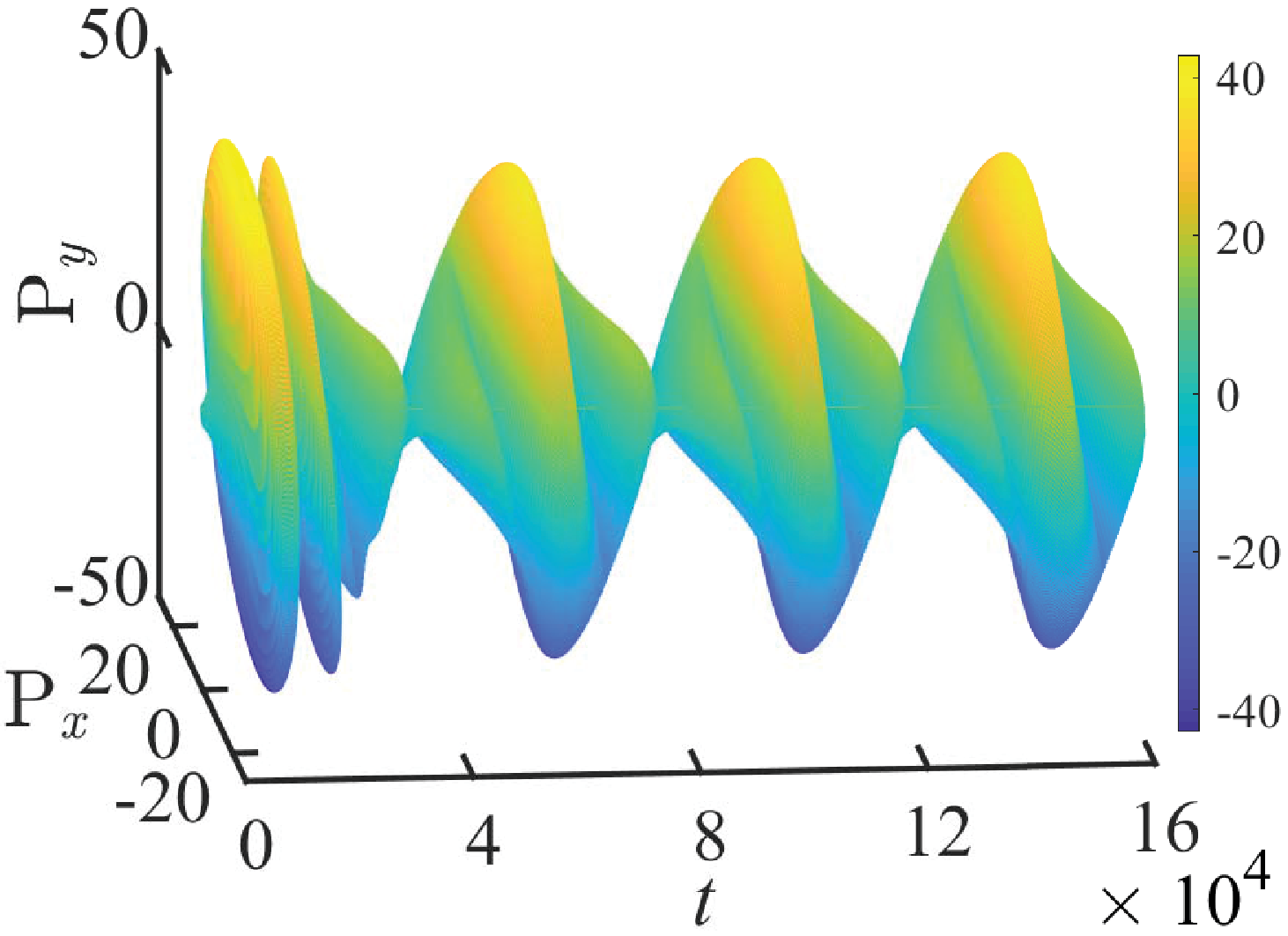} &
\includegraphics[width=0.4\textwidth]{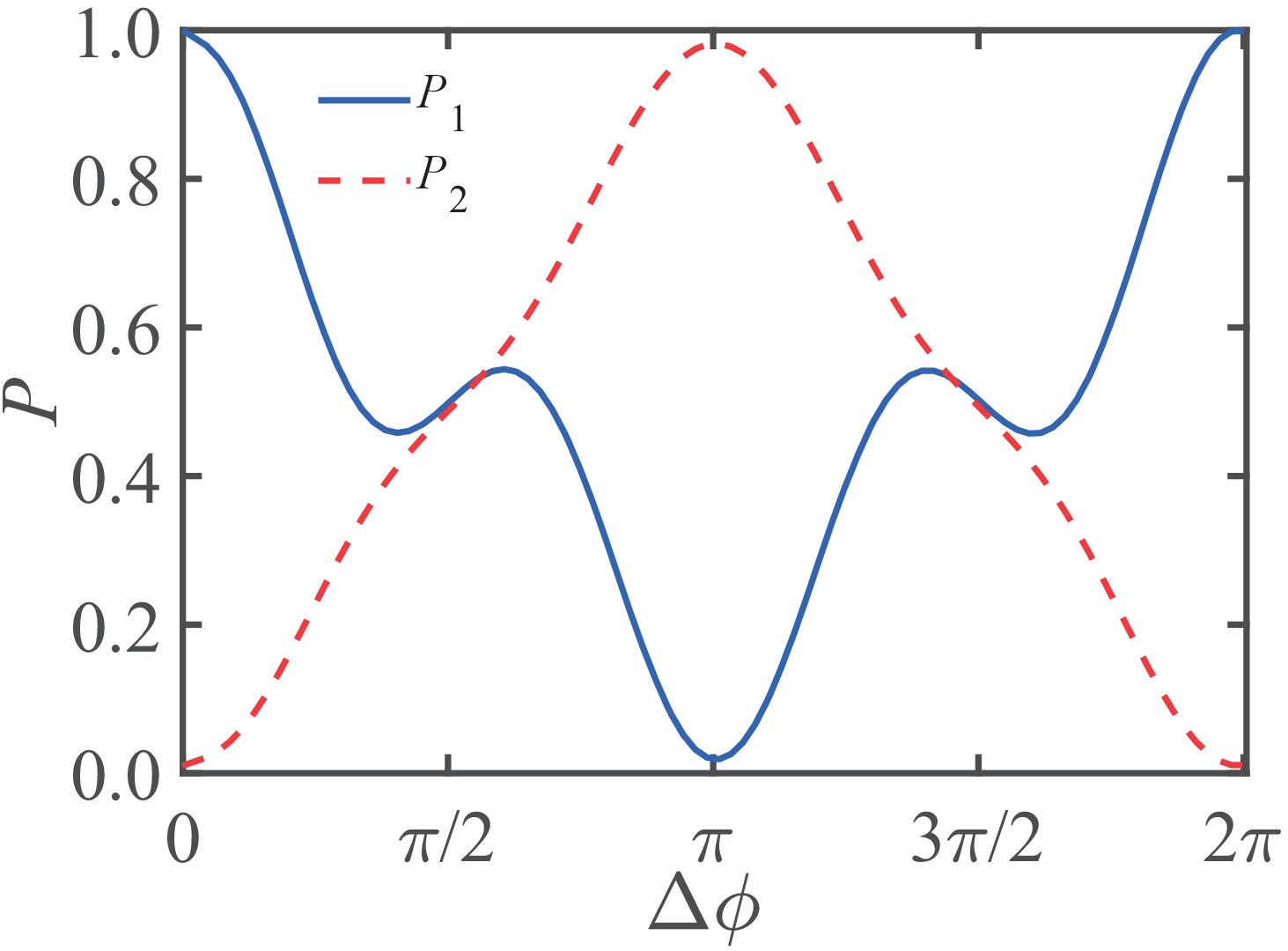} \\
(c) & (d)
\end{tabular}
\protect\caption{Vorticity driven VB oscillation explored in detail. {(a) Periodic cell drifting in coordinate space for $u(x)$.} (b) Vortex like dynamics in a two dimensional subspace. {(c) The time-evolution of state vector onto the an initially unstable subspace.} (d) The overlapping of the 4D degenerate subspace [$\vec{e}_{3\sigma}$ $\vec{e}_{4\sigma}$], as in Eq.~\cmmnt{(\ref{lc0})}(25), with the corresponding unstable (solid blue line) and stable (dotted red line) eigenmodes of the full $N\times N$ matrix $(\vec{D}+\vec{Q})\vec{R}$.}\label{SFig6}
\end{figure}

In a VB mode, when an unstable vortex like circulation increases in size, eventually this growth ceases because of the nonlinearity and decays at some later time. We find that, simultaneously, the periodic cellular structure itself undergoes phase drifting, indicating that this is driven by the vorticity of the former. In fact, closer examination shows that, as the cells drift, the overlap of the growing eigenstate with a shrinking eigenstate of the drifted stationary state increases. When the drift phase reaches $\pi$, the overlap is a maximum. Clearly, the shrinking mode causes the vortex circulation to return to zero. Note we only look at one pair of modes. There are other modes acting so that the drifting continues and forms a limit cycle when the phase reaches $2\pi$. The complete dynamics is shown in Fig.~\ref{SFig6}. The system is initially in the $3^{\rm rd}$ subspace and evolves by Eq.~\cmmnt{(\ref{lc0})}(25) with $L = 512$, $h = 0.32$, $\alpha = 0.12$, and $\kappa = 0.6381$. The projection on to the subspace changes with time {, ${\rm{P}}_{\sigma}(t) = \vec{e}^{\dag}_{3\sigma}\cdot\tilde{\vec{u}}(t)$ ($\sigma=x,y$ for convenience)}. From the perspective of the cross section, this is a vortex like motion in time with the time-step set to ${\rm{d}}t = 0.003$, and data is recorded every 400 iterations. The saturation is clearly present when it reaches a certain magnitude and the motion becomes quasi limit cycle. As mentioned above, this happens while cellular structure itself undergoes phase drifting in coordinate space. The behavior can be understood from the overlap, $P$, between the 4 dimensional degenerate subspace [$\vec{e}_{3\sigma}$ $\vec{e}_{4\sigma}$] with the stable and unstable eigenmodes of the complete $(\vec{D}+\vec{Q})\vec{R}$ matrix calculated at respective stationary points with phase drift $\Delta\phi$. It is found that the overlaps between stable and unstable oscillate periodically with $\Delta\phi$.

\section{Data Availability}
The necessary formulae and the steps to perform the computations are detailed in the main work and the Supplementary Information. There is no need for a specific platform nor a specially purposed software package. The data used to justify the results and conclusions of this work are entirely presented within the body and supplementary information of the manuscript.

The code used in this paper is available on request from chenyongcong@shu.edu.cn.

\clearpage
